\documentclass[bibyear]{aa}
\usepackage{graphicx}
\usepackage{lipsum}
\usepackage{natbib}
\usepackage{amsmath}
\usepackage{multicol}
\usepackage{stfloats}
\usepackage{color}
\usepackage[flushleft]{threeparttable}
\usepackage{yfonts}
\usepackage{ulem}
\usepackage[dvipsnames]{xcolor}
\usepackage{txfonts}
\usepackage{wedn}
\usepackage[T1]{fontenc}
\usepackage[utf8]{inputenc}
\usepackage{url}
\usepackage{hyperref}

\hypersetup{
    colorlinks=true,
    citecolor = blue,
    linkcolor=blue
            }

\usepackage{yfonts}
\usepackage{xcolor}

\usepackage{subfigure}
\newcommand{\teff}{$T_{\rm eff}$\,}
\newcommand{\degree}{$^\circ$\,}
\newcommand{\kms}{$\mathrm{km\,s}^{-1}$}

\usepackage{aurical}
\usepackage[T1]{fontenc}
\usepackage{pdfcomment}

\begin{document}

    \title{Magnetic activity on the young Sun: a case study of EK\,Draconis}

   \author{A.~Görgei
          \inst{1,2,3}
          \and 
          L.~Kriskovics
          \inst{1,2}
          \and
          K.~Vida
          \inst{1,2} 
          \and
          B.~Seli
          \inst{1,2,3}
          \and
          K.~Oláh
          \inst{1,2}
          \and
          P.~Sági
          \inst{1,2,3}
          \and
          A.~Bódi
          \inst{1,2,4}
          \and
          S.P.~Järvinen
          \inst{5}
          \and
          K.G.~Strassmeier
          \inst{5,6}
          \and
          A.~Pál
          \inst{1,2}
          \and
          Zs.~K\H{o}v\'ari
          \inst{1,2}
          } 

   \institute{Konkoly Observatory, HUN-REN Research Centre for Astronomy and Earth Sciences, Konkoly Thege \'ut 15-17., H-1121 Budapest, Hungary\\
              \email{gorgei.anna@csfk.org}
        \and
            HUN-REN CSFK, MTA Centre of Excellence, Konkoly Thege \'ut 15-17., H-1121 Budapest, Hungary
         \and
             E\"otv\"os Lor\'and University, Institute of Physics and Astronomy, Department of Astronomy, P\'azm\'any P\'eter s\'et\'any 1/A, H-1117 Budapest, Hungary
        \and
            Department of Astrophysical Sciences, Princeton University, 4 Ivy Lane, Princeton, NJ 08544, USA
        \and
             Leibniz-Institut für Astrophysik Potsdam (AIP), An der Sternwarte 16, 14482, Potsdam, Germany
        \and
             Institut für Physik und Astronomie, Universität Potsdam, 14476, Potsdam, Germany
             }

   \date{Received ...; accepted ...}

  \abstract
   {Young, solar analogue stars provide key insights into the early stages of stellar evolution, particularly in terms of magnetic activity and rotation. Their rapid rotation, high flaring rate, and enhanced surface activity make them ideal laboratories for testing stellar models or even the solar dynamo.} 
   {Using long-term photometric data, we investigated the cyclic behaviour of EK\,Dra over the last century. We analyze its short-term activity based on 13 Transiting Exoplanet Survey Satellite (TESS) sectors. Applying Doppler imaging on high-resolution spectral data we investigate short and long-term spot evolution and surface differential rotation.}
   {We use Short-term Fourier-transform on a 120 years long archival photometric data in order to search for activity cycles. The short-term space photometry data is fitted with an analytic three-spot model, and we hand-select flares from it to analyze their phase and frequency distribution. Spectral synthesis is used to determine the astrophysical parameters of EK\,Dra. Using the \texttt{iMap} multi-line Doppler imaging code, we reconstruct 13 Doppler images. Differential rotation is derived by cross-correlating consecutive Doppler maps.}
   {Long-term photometric data reveal a 10.7-12.1 year cycle that was persistently present for $120$ years. In the more recent half of the light curve a 7.3-8.2 years-long signal is also visible. The distribution of the 142 flares in the TESS data shows no correlation with the rotational phase or with the spotted longitudes. The reconstructed Doppler images show a surface that varies from rotation to rotation, putting the lower limit of the spot lifetime between 10-15 days. Based on the cross-correlation of the Doppler maps, EK\,Dra has a solar-type differential rotation with a surface shear parameter of $\alpha_{\rm DR}=0.030 \pm 0.008$.}
   {}

   \keywords{stars: activity --
                stars: imaging  --
                starspots --
                stars: individual: EK\,Dra
               }

   \maketitle

\section{Introduction}

Understanding the life cycle of stars like our Sun is essential for unraveling the physical processes that govern stellar evolution and the environments of planetary systems. Young, solar analogue stars provide key insights into the early stages of stellar evolution, particularly in terms of magnetic activity and rotation \citep{Vidottoetal2014}. These early stages are often characterized by high-energy phenomena such as flares, strong stellar winds, and increased spot activity \citep[e.g.,][]{Gudel2004,Vidaetal2024}. Such behaviour can have profound effects on the atmospheres and habitability of surrounding planets \citep[][etc.]{Lammeretal2003,Ribasetal2005,Vidaetal2017,Vidaetal2019}. 

By studying stars at different stages of their evolution, we can reconstruct the past of the Sun and predict its future. Of particular interest among young solar analogues are stars that are very similar to the Sun in mass and chemical composition, but are significantly younger. These stars allow us to observe solar-like magnetic dynamo processes in a more vigorous and dynamic state. Their rapid rotation, high flaring rate, and enhanced surface activity make them ideal laboratories for testing solar and stellar models. Accordingly, the focus of this paper is on one such star – EK\,Draconis (HD\,129333), a young, single and rapidly rotating G-type dwarf whose extreme activity and solar-like properties make it a key target for understanding solar-stellar magnetism, rotation, and not least the impact of all of this on planetary environments \citep[cf.][]{Namekataetal2022a,Namekataetal2022b}.

EK\,Dra, partly because of its high brightness, has been observed in various surveys of late-type stars targeting magnetic activity well before the discovery of its variability. Two of the early attempts are (i) \citet{1985AJ.....90.2103S} observed chromospheric emission connected with stellar rotation, and (ii) \citet{1989A&A...213..226C} searched for activity signatures in the IUE archives comparing the results to SKYLAB spectra made on different parts of the solar surface; both surveys published observations of EK\,Dra. The star is very similar to the Sun in almost every parameter except that it rotates ten times faster, and the consequence of this, due to the rotation-activity relation \citep[e.g.,][]{1967ApJ...150..551K,1972ApJ...171..565S,1984ApJ...279..763N}, is its very high-level magnetic activity. EK\,Dra is the usual G-dwarf example in hundreds of papers studying active stars, and has been a part of many further surveys ever since. Moreover, EK\,Dra has a faint, low-mass companion star (EK\,Dra~B) in a wide binary system with an orbital period of $\sim$45 years and a distance of 2.2\,AU at periastron \citep{2005A&A...435..215K}, that is, there is unlikely to be any significant interaction between the two components that could affect the magnetic activity of EK\,Dra~A.

The discovery of the light variation of EK\,Dra due to starspots was made by \cite{1991IBVS.3680....1C}. By that time several different observations had already existed and been published that pointed to magnetic activity as the origin of the variability. Direct measurements of the magnetic field were published by \cite{2020A&A...635A.142K}. The history of photometric observations is well described in \cite{2017MNRAS.465.2076W}. The summary of all previous attempts at Doppler imaging is found in \cite{jarvinen2018}, where the photometric cycle of EK\,Dra is also shown until 2018. 

This study aims to investigate the short- and long-term activity-related phenomena (cycles and flares) using archival light curves. We also aim to expand the existing pool of Doppler images with multiple consecutive maps, with the goal of making the short-term changes in surface structure visible and deriving surface differential rotation.

\section{Observations}\label{sect:obs}

\subsection{Photometric observations} \label{section:phot_obs}

We used photometric observations from multiple different sources. Transiting Exoplanet Survey Satellite (TESS, \citealt{tess}) provides high-cadence light curves, which are suitable for investigating short-term brightness changes (e.g. effects of spots and flares). 
Digital Access to a Sky Century @ Harvard (DASCH) archive has multiple decades worth of data from scanned photographic plates, which provides the opportunity to observe the long-term behaviour of EK\,Dra. To extend the time-frame covered by DASCH, we use data from two publications: \cite{BminV} and \cite{jarvinen2018}, collected with the Automated Photoelectric Telescopes (APTs) at Fairborn Observatory, Arizona.

TESS was launched in April 2018 and since then has been on a 13.7 day long orbit around Earth. 
It is equipped with four 10.5\,cm telescopes and covers almost the entire sky in 24\degree $\times$ 90\degree instantaneous strips (sectors). 
One sector provides $\sim$27\,d long continuous observations with short gaps that are the result of data transfer to Earth. 
EK\,Dra was observed in 13 Sectors: S14, S15, S16, S21, S22, S23, S41, S48, S49, S50, S75, S76, and S77. We used the 120\,s cadence Pre-search Data Conditioning Simple Aperture Photometry (\mbox{PDCSAP}) light curves provided by the Science Processing Operations Center \citep[SPOC;][]{10.1117/12.2233418} for flare search, except for Sector 50 where it is not available. We used the \texttt{fitsh}\footnote{\url{https://fitsh.net}} \citep{fitsh} based \texttt{qdlp-extract} pipeline to reduce full-frame images for spot modeling. The first provides better time resolution, while the latter yields light curves with fewer gaps.

The DASCH project digitizes the Harvard College Observatory Astronomical Photographic Glass Plate Collection for scientific applications covering roughly 100-year time scales. \citep{dasch0,dasch1,dasch2,dasch3}. 
The data were recorded on photographic plates that have a color response close to Johnson $B$ without any filter and yield $\sim$0.1 mag photometric scatter \citep{dasch3}; their digitization was finished in March 2024. The first measurement of EK\,Dra is from 1891 and the last from 1989, providing almost a century-long light curve. 
  
This light curve was extended by the data from \cite{BminV} and \cite{jarvinen2018}. \cite{BminV} combined the V-band observations of three APTs operated at the Fairborn Observatory in southern Arizona. In \cite{jarvinen2018} the observing instrument was the T7 Amadeus telescope, one of the two 0.75m Vienna-AIP APTs \citep{kgs:boyd97} which also recorded $V$-band data.

\subsection{Spectroscopic observations} 

Spectroscopic observations for Doppler imaging were obtained with the fiber-fed ACE \'echelle spectrograph ($R$=21000) mounted on the 1.02m Ritchey-Chr\'etien-Coudé telescope at Piszk\'estet\H{o} Mountain Station of Konkoly Observatory, Hungary. The covered wavelength range of the instrument is 4200$-$8500\,\AA\, with a large number of suitable Doppler imaging lines.

The spectra were obtained on a total of 90 nights between February 2021 and July 2024. Since the rotational period of the star is $\sim$2.6 days, the minimum time needed to cover the phases for Doppler imaging is a consecutive 5 nights of observations. The exposure times were 15-30 minutes depending on the seeing, but never longer to avoid rotational smearing affecting the results. During the imaging process, we prioritized using spectra with a signal-to-noise ratio (SNR) greater than 100. The only exceptions to this are a few cases where omitting low-quality data would have resulted in a considerably worse phase coverage. In this way, we were able to create subsets for 13 rotations. On two occasions, three rotations are within a three-week period, but otherwise the spacing of the observations is larger. The observations were phased with the following ephemeris from \cite{jarvinen2018}:

\begin{equation}
    \label{eq1}
\hspace{0.25\linewidth}
HJD = 2445781.859 + 2.606 \times E.
\end{equation}

The observational data were reduced by \texttt{piszkespipe}\footnote{\url{https://pypi.org/project/piszkespipe}}, a modification of CERES \citep{piszkespipe}. This employs the standard reduction techniques of bias subtraction, flat-field correction to remove pixel-to-pixel variations and the curvature of the blaze function, correction for the dark current, extraction of the \'echelle orders and wavelength calibration with the spectra of a thorium-argon lamp.

\section{Photometric analysis}

\subsection{Long-term behavior}\label{section:long-term-phot}

The long-term behaviour of EK Dra was investigated based on the folűlowing datasets: DASCH data between JDs $2411786.89$ and $2445820.95$, \cite{jarvinen2018} APT data between JDs $2451669.88$ and $2458237.89$, and \cite{BminV} data were used to bridge the gap between. \cite{BminV} overlaps with both DASCH and \cite{jarvinen2018} data, since it includes measurements between JDs 2445806.90 and 2451675.79. We note that this is a dataset that consists of data obtained from four different sources, where each data point is an average of minimum five, maximum 66 days of observations. In addition, DASCH data were averaged every 20 days. The DASCH and APT light curves are measured in different filters. To make them comparable for the period analysis, we apply an approximate correction for the rotational amplitudes. We combine BT-NextGen \citep{nextgen} model spectra assuming $T_\mathrm{eff}=5700$\,K for the photosphere, $T_\mathrm{eff}=4600$\,K for the starspots, and a spot filling factor of 0.2 (see Sect.~\ref{sect:doppi}). Using the Johnson $B$ and $V$ transmission curves we find an amplitude ratio of 0.88. We scale the $V$-band APT light curve with this factor, and shift it to the DASCH dataset using the median magnitude in the overlapping segment. The resulting light curve was searched for periodic variations using the Short-Term Fourier-Transform (STFT) method as implemented by \cite{tifran}. 

\begin{figure}
\includegraphics[width=\columnwidth]{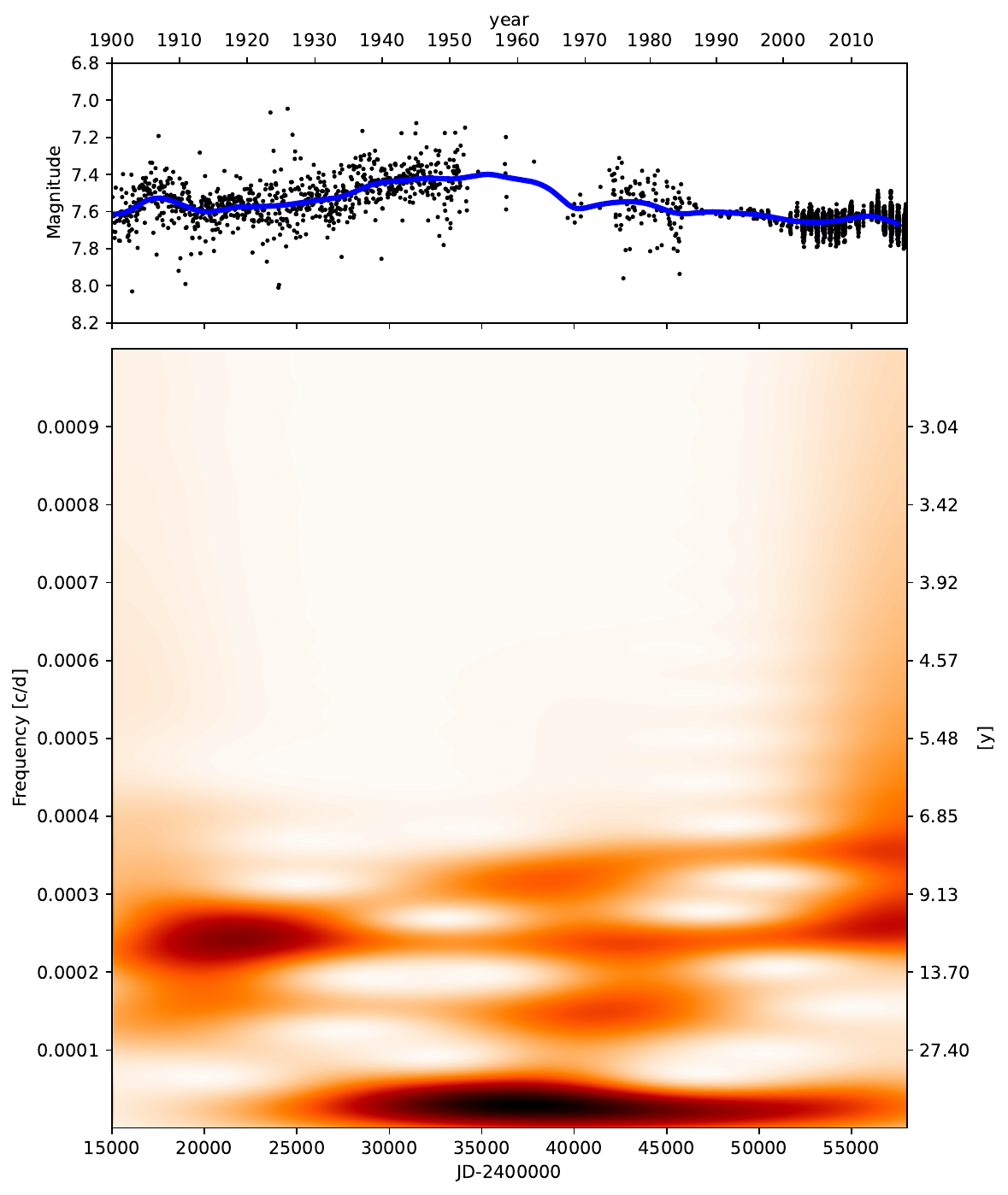}
\caption{Short-term Fourier transform of the dataset used in Sect.\,\ref{section:long-term-phot}. A 10.7-12.1 years long cycle is seen through the whole plot, and in the more recent half of the light curve a 7.3-8.2 year-long period also appears. Features longer than 27 years are suppressed on the plot to 1/3 power. The peak in this range originates from the length of the available data and the gap between 1960-1970.}
\label{fig:stft}
\end{figure}

Figure\,\ref{fig:stft} shows the combined light curve and the result of the STFT.  Throughout the observed time, a 10.7-12.1 years long cycle is seen. In the more recent half of the light curve a 7.3-8.2 years long signal is also visible. 
We conducted a false alarm probability analysis for all STFT windows. Assuming that there is no periodic signal in the data, we would observe these peaks 0.1\% of the time, which is a strong indication that the observed periodicity is present in the data. In the beginning of the light curve the star shows a brightening phase, then from JD $\sim$2435000, it shows a fading trend over the past $\sim$64 years. This, whether part of a cycle or a trend, certainly points to centuries of change.

\subsection{Analytic spot model}\label{sect:3spot}

We fitted the light curve obtained from the long-cadence full-frame TESS images with an analytic three-spot model following the equations of \cite{3spot_eq} as implemented by \cite{py_3spot}. The method uses circular spots that are allowed to overlap with each other. Since the light curves show clear signs of spot evolution, we applied time-series spot modeling: the time interval is divided into sub-intervals and for each sub-interval, a regular analytic model is fitted, resulting in a set of spot parameters which are a function of time. The width of the time series window was 1.25\,$P_{\rm rot}$ and the step size was 0.25\,$P_{\rm rot}$ where the rotational period was kept at $P_{\rm rot}=2.606$\,d for comparison with the Doppler maps. In the overlapping parts, final theoretical brightnesses are obtained by averaging the theoretical brightnesses of each overlapping part \citep{1996KOTN....6....1B}. The Levenberg--Marquardt algorithm was applied during minimization.

 Photometric starspot modeling is an inherently degenerate problem: out of the four parameters, which describes an analytic spot (i.e., spot latitude and longitude, spot temperature and spot radius), only the longitude is reliably recoverable. The other three parameters are not independent, e.g. a larger change in the light curve can be caused by a cooler spot, a larger spot, or a different spot latitude depending on the inclination. Hence, we chose a simplified modeling approach. During the fitting process, the spot latitudes were kept at 30\degree and the spot temperatures at 4600\,K. These results are based the Doppler map of \cite{jarvinen2018}. 
 Their umbral spot temperature difference for their most dominant spot is $\Delta T$$\approx$1000\,K which we applied to our rounded effective temperature derived from spectral synthesis (see Table \ref{tab:params}). We note that we chose not to use our Doppler maps due to the fact that spot temperatures can be distorted by the smearing effect of the lower resolution (see Appendix \ref{appA} and \citealt{2023A&A...674A.143K}). The spot longitudes and the sizes of the spots were the free parameters of the fits. The rotational period used during the light curve fitting was kept at the same constant value used for the Doppler inversions.  

Our seasonal period analysis shows clear signs of changes hinting at either or both differential rotation and changing spot latitudes. However, when all spot parameters are fitted, the degeneracy of the problem often renders the interpretation and comparison of the spot parameters unreliable. Our approach makes the comparison between Doppler maps and the photometric models clearer, and it also enables us to carry out the analysis described in Section \ref{subsect32}, which were our primary goals with the photometric spot modeling.

The fitting was carried out for the following sectors: S14, S15, S16, S21, S22, S23, S41, S48, S49, S50, S75, S76, S77. The results are shown on Fig.\,\ref{fig:sector14} and Fig.\,\ref{fig:S15-22}-\ref{fig:S75-77}. The top of each three-panel figure shows the light curve from the full-frame images in gray, and the three-spot model fits in blue. 
The middle panel denotes the spot longitudes for each fitting window.
Here, the sizes of the dots are scaled by the spot radii, and in the upper left corners an ${r}$=10\degree circle is shown for reference. At the bottom the PDCSAP light curve is visible along with the flares marked in red. The vertical lines indicate the times when the spots are in the line of sight. 

\subsection{Change in the spot longitudes}\label{subsect32}

\begin{figure*}
\includegraphics[width=2\columnwidth]{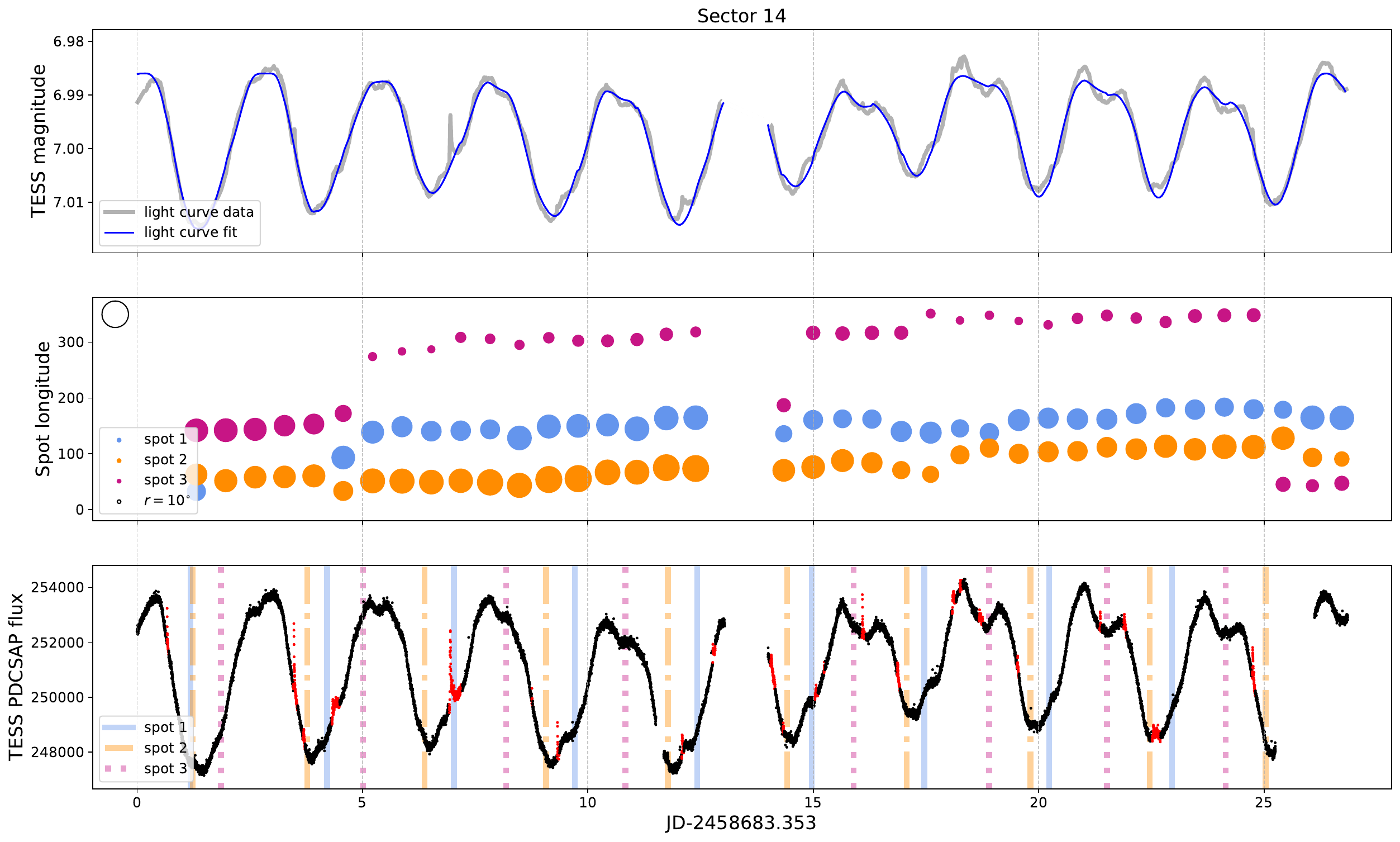}
\caption{TESS light curve of EK\,Dra from Sector 14 with the spot model. Top: The light curve from the full-frame images (gray) and the 3-spot-model fit (blue). Middle: The change in spot longitudes. The sizes of the dots are scaled by the spot radii. 
Bottom: The PDCSAP light curve along with the flares marked in red. Vertical lines indicate the times when the spots are in the line of sight.
}
\label{fig:sector14}
\end{figure*}

The shift of the spot longitudes in the middle panels of Fig.\,\ref{fig:sector14} and Fig.\,\ref{fig:S15-22}--\ref{fig:S75-77} can be used to infer the operation of differential rotation. During a suitably chosen short interval the longitude values of a spot with a rotation period different from the photometric period change with some slope compared to the horizontal. Based on the TESS light curves, it appears that each such interval is about 10-15 days, or roughly half a TESS sector. Within these intervals, the spots are considered stable, that is, spot evolution is disregarded. (We note here that all this is consistent with the Doppler images to be presented in Sect.\,\ref{sect:doppi}.) The changes in spot longitude from photometric spot modeling within each time interval are approximated by a linear function, and by comparing the different slopes we can see whether the spots rotate faster or slower compared to the photometric period. The average of the fitted slopes is 1.3$^\circ$/d, and the standard deviation is 4.6$^\circ$/d. This spread in the slopes hints to differential rotation. However, a more thorough determination is not possible with this method. We would only be able to estimate the magnitude of the differential rotation if we knew the exact latitude coordinates of the spots, which can only be derived from photometric spot models with great uncertainty.

\begin{figure}
\includegraphics[width=\columnwidth]{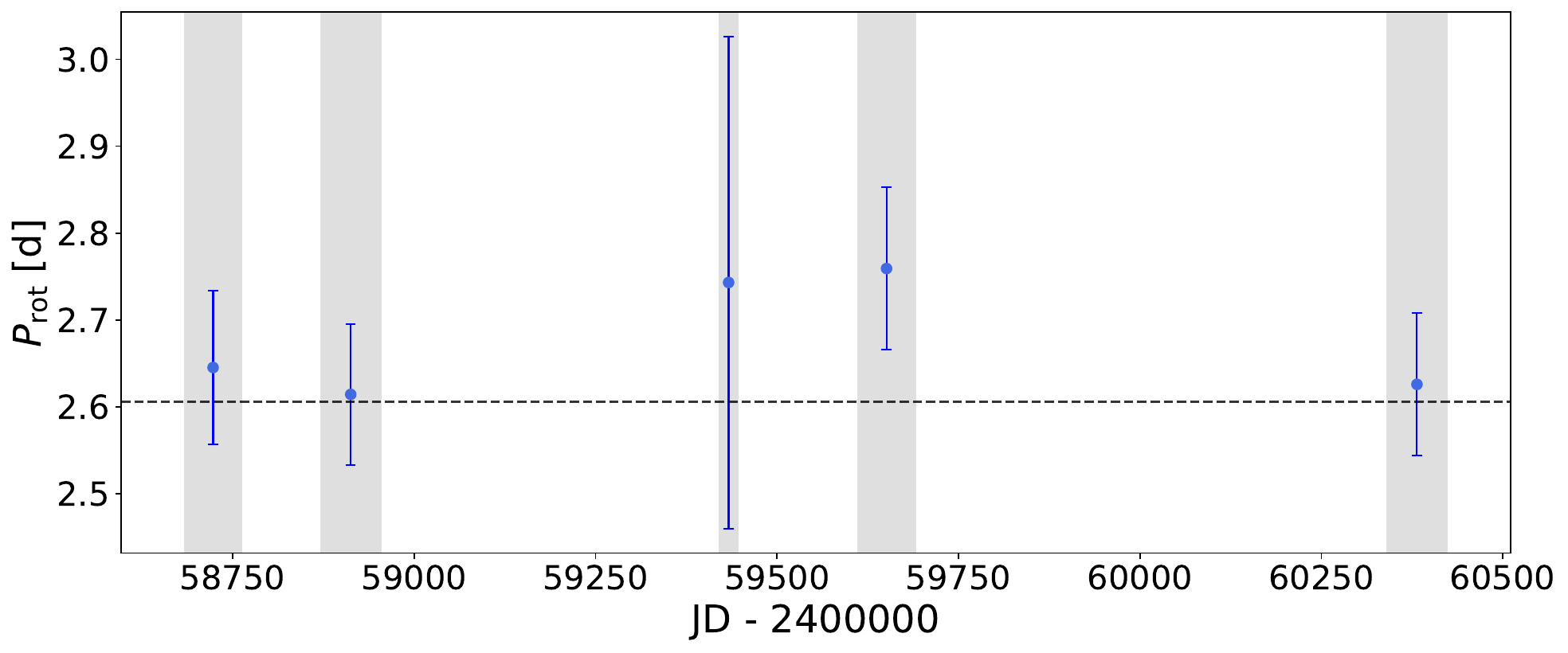}
\caption{Rotational periods of EK\,Dra (blue dots) and their estimated error bars in consecutive TESS observation windows (adjacent sectors are considered continuous). Gray areas from left to right are sectors S14-16, S21-23, S41, S48-50, and S75-77. The horizontal dashed line marks the rotation period value adopted for Doppler imaging.}
\label{fig:prot_change}
\end{figure}

\subsection{Change of the rotational period due to differential rotation}

We determined the rotational period for consecutive TESS sectors (S14-16, S21-23, S41, S48-50, and S75-77) using windowed Lomb-Scargle transform as implemented by \cite{Bodi2024}. The period uncertainty $\delta P$ is calculated for the continuous observational windows from their obtained rotational period $P_{\rm rot}$ and duration $T_{\rm N}$ according to the relation $\delta P\approx P_{\rm rot}/T_{\rm N}$. Figure\,\ref{fig:prot_change} shows the results obtained for each observation period. The average relative deviation of the rotational periods calculated from these is $\Delta P/\overline P_{\mathrm{rot}}=0.02 \pm 0.02$, which can be a lower estimate of the surface differential rotation (relative shear), if we assume that the cause of the period change is primarily the differential rotation and the change in the latitudinal positions of the dominant spots. For previous applications of this photometric method, see e.g. \citet{2002A&A...394..505B,2017AN....338..453S,2021A&A...650A.158K,2025A&A...701A.103K}.

\subsection{Flares}
\label{sect:flares}

We investigate the flaring characteristics of EK\,Dra based on the PDCSAP light curves in  the available 12 TESS sectors. The flares were collected by visual inspection of the PDCSAP light curves one by one. We found 142 flares this way. We note that the noise level of the light curve changes smoothly for most sectors, except for S76 where it increased abruptly. This hinders the detection of smaller flares in that sector, but as we are not using a flare finding algorithm that is directly tied to the photometric scatter, we estimate that its impact on the flare statistics is negligible.

\begin{figure}
\includegraphics[width=\linewidth]{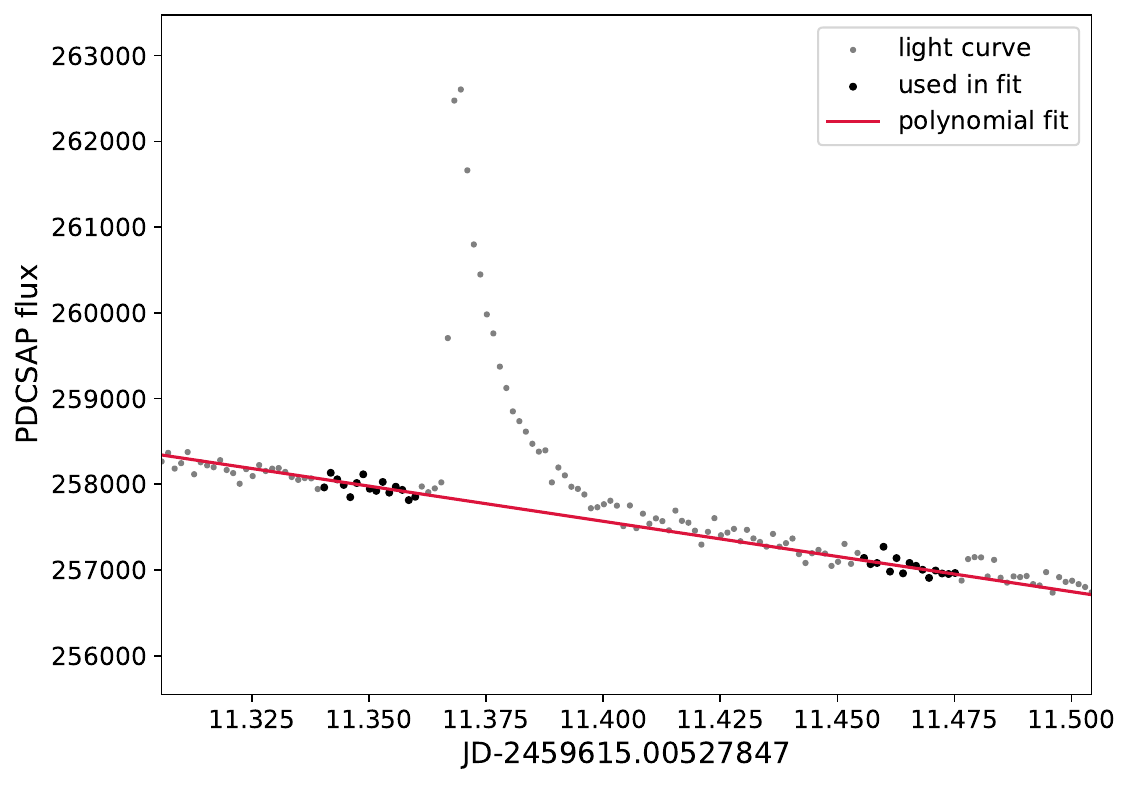}
\caption{Example of the polynomial fitting of the baseline of the flares. The data is marked gray, the black points were used for the fit, and red shows the fitted baseline. The flare in the figure is from Sector 48.}
\label{fig:flare}
\end{figure}

In order to calculate the flare energies, we used the same approach as in \cite{flareEmethod1} and \cite{flareEmethod2} described in the following. After marking the flaring data points and normalizing the light curves, we first fitted fourth-order polynomials to the 30$-$30 minute light curve segments bordering each individual flare on both sides (Fig.\,\ref{fig:flare}). The baseline defined this way was removed as the next step. On the resulting data, we calculated the area below each flare (equivalent duration, ED) by integrating according to Simpson's rule\footnote{\texttt{scipy.integrate.simps}}. 
We integrated a BT-NextGen model spectrum \citep{nextgen} with $T_{\text{eff}}$=5700\,K, $\log g$=4.4 and solar metallicity over the entire wavelength range with and without convolving with the TESS response function, to calculate the ratio of TESS ($L_{\text{TESS}}$) to bolometric luminosity ($L_{\text{bol}}$). By applying the bolometric correction on the absolute magnitude using the method described in \cite{gaia_bolcorr} and the data from \cite{gaia_dr3}, the value of $L_{\text{bol}}$ was determined.
With the above considerations, the flare energy in the TESS band is calculated as follows:
\begin{equation}
\hspace{0.35\linewidth}
E_{\text{TESS}} = L_{\text{TESS}} \times ED.\quad 
\end{equation}

\begin{figure}
\includegraphics[width=\linewidth]{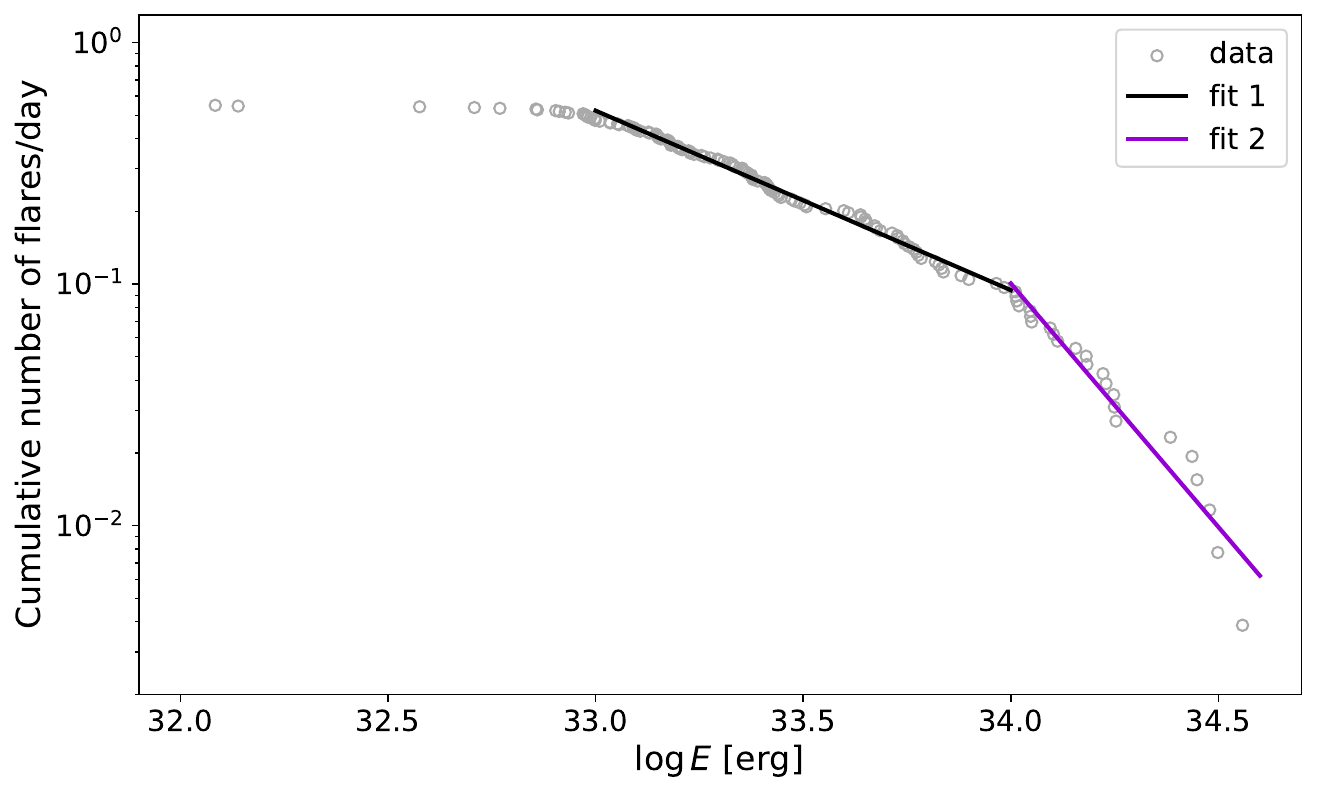}
\caption{The flare frequency distribution from TESS PDCSAP data. The black line denotes the fit with a power-law index of $ 1.466 \pm 0.007$ in the 10$^{33}$--10$^{34}$\,erg energy range. The purple line is a fit above 10$^{34}$\,erg, its power-law index is $2.335\pm 0.110$.}
\label{fig:ffd}
\end{figure}

The flare frequency distribution diagram (FFD) is shown in Fig.\,\ref{fig:ffd}. At low energies the flare events have a lower recovery rate due to the light curve noise, so only flares with energies higher than $ E=10^{33}$\,erg were considered in the following analysis. (For a detailed flare injection test, see \citealt{2020A&A...641A..83K}.) A breaking point around $E=10^{34}$\,erg divides the FFD into two parts. The fit between $E=10^{33}$\,erg and $10^{34}$\,erg yields the power law index of $1.466 \pm 0.007$ while the one above $E=10^{34}$\,erg is $ 2.335\pm 0.110$.

We investigated the longitudinal distribution of flares by applying Kuiper's test \citep{kuiper1960tests} to our data to determine whether
it comes from a uniform phase distribution. Unlike the similar Kolmogorov$-$Smirnov test, it is invariant under cyclic transformations, making it suitable for investigating the cyclic phase distribution. Based on this test, we cannot reject the null hypothesis of a uniform flare phase distribution. However, as the starspots of EK\,Dra evolve quickly, it might be more suitable to compare the phase preference of flares to the spot locations.

To investigate whether the flares are associated with the active regions responsible for the spot modulation observed on the TESS light curves, we propose a simple test. Using the results of the analytic three-spot model described in Sect.\,\ref{sect:3spot}, we assign each flare to the spot closest in longitude. We assume that the flare originates from the meridian, thus we can use the phase of the flare mid-time to calculate its longitude. We then calculate the longitudinal difference between the flare and the closest spot, and compare these differences to artificial distributions, where the same number of flares are distributed randomly in time. We use a two-sample Kolmogorov$-$Smirnov test to compare the observed distribution to the artificial ones. Running 1000 tests with different random flare times, we cannot reject the null hypothesis of random flare phase occurrence (with median $p$-value of 0.5). However, since there are visible spots at every rotational phase, the assumption that the flares originate from the meridian is likely violated, meaning that this simple test can only provide weak evidence for random flare occurrence.

\section{Spectroscopic analysis}
\subsection{Age}

\cite{Li_age} found strong correlation between the age of young solar analogue stars and their lithium abundances. Using this empirical correlation and the lithium abundance from the result of our spectral synthesis (NLTE corrected value: $A$(Li)=2.28$\pm$0.04), the age of EK\,Dra is estimated to be  $t$=0.070$\pm$0.018\,Gyr. 
This reinforces the presumption that EK\,Dra is a young solar analogue and is in good agreement with previous results \citep[e.g.][]{2007LRSP....4....3G}.

\subsection{Doppler imaging}\label{sect:doppi}
    
\subsubsection{Astrophysical parameters}
      
Precise astrophysical parameters are required for successfully reconstructing the surface of a star through Doppler imaging. For this reason, we performed a spectroscopic analysis based on spectral synthesis using Spectroscopy Made Easy (SME) \citep{piskunov_sme} with MARCS atmospheric models \citep{gustafsson_marcs} and atomic line parameters taken from the Vienna Atomic Line Database \citep[VALD,][]{kupka_vald}. The value of the macroturbulence was estimated following the relation from \citet{valenti_macro}:

\begin{equation}
\hspace{0.2\linewidth}
    v_{\mathrm{mac}}=\Bigg(3.98-\frac{T_{\mathrm{eff}}-5770 \mathrm{K}}{650 \mathrm{K}}\Bigg)\,\mathrm{km\,s}^{-1}.
    \label{macro_eq}
\end{equation}
    
The final values of the astrophysical parameters were determined during an iterative process. The steps were the following:

 \begin{itemize}
    \item{The initial astrophysical parameters were taken from \cite{jarvinen2018}}
    \item{$\log\,g$ was fitted using stronger lines ($\log gf > 0$)}.
    \item{The effective temperature was fitted, using $\log g$ from the previous step.}
    \item{Fitting the metallicity with the value of $\log g$ and $T_{\text{eff}}$ being kept constant.}
    \item{Fitting of $v\sin i$ and $v_{mic}$.}
    \item{Refitting $\log\,g$, $T_{\text{eff}}$ and metallicity simultaneously to confirm robustness.}
 \end{itemize}
 
The results are summarized in Table\,\ref{tab:params} along with the other stellar parameters taken from the literature. These values were used for all of the Doppler inversions.
 
\begin{table}
\caption{Stellar parameters of EK\,Dra}
\centering
\begin{tabular}{lll}
\hline\hline
Parameter & Value & Source\\
\hline
Sp. type & G1.5V & \cite{strassmeier_rice_1998} \\
Li age & $ 0.70 \pm 0.18 $  Gyr & this paper\\
$P_\mathrm{rot}$ & $2.606$\,d & \cite{jarvinen2018} \\
$v \sin i$ & 17.5 \kms & \cite{jarvinen2018}\\
$i$ & $63 \pm2$\degree &\cite{jarvinen2018}\\ 
\teff &$5682 \pm 75$ K& this paper\\
$\log g $ &$4.46 \pm 0.22$ & this paper\\
Metallicity &$-0.20 \pm 0.03$ & this paper\\
$\nu_\mathrm{mic}$ & $2.49\pm0.35$\,\kms & this paper\\
$\nu_\mathrm{mac}$ & $3.84$ \kms & this paper\\
\hline\\
\end{tabular}
\label{tab:params}
\end{table}

\subsubsection{The Doppler imaging code \texttt{iMap}}

Doppler imaging was carried out with \texttt{iMap} \citep{carroll_imap}.
It performs multi-line Doppler inversion using a list of user-provided photospheric lines. These relatively unblended lines were chosen from the 5000-7000\,\AA{} range and have a well-defined continuum and suitable temperature sensitivity. The lines in this list are referred to as "imaging lines". The code uses a full radiative solver \citep{carroll_solver} for every local line profile on the stellar surface that was divided into $5^{\circ}\times5^{\circ}$ segments.
Atomic line data for the Doppler inversion were taken from VALD. The model atmosphere from \cite{castelli_mod} was interpolated for the required temperature, gravity, and metallicity. Instead of a spherical model atmosphere, the code uses LTE radiative transfer, where the multiline approach compensates for the imperfections in the fitted line shapes. The stellar parameters used for the inversion are listed in Table \ref{tab:params}.
Finally we note, that during image reconstruction, no extra constraints were added, as surface reconstruction uses an iterative regularization method based on the Landweber algorithm \citep[see][]{carroll_imap}.

\subsubsection{Surface reconstruction and temporal variations}
   
We observed more than 900 spectra during 30 months. These were divided into 13 subsets (DI01-DI13), then from each of these subsets, we selected the ones with the highest SNR to cover the phases of the rotation. (Detailed information on the subsets can be found in Table\,\ref{tab:DI_info} and Fig.\,\ref{fig:phases_for_di}). In some cases, in order to minimize the phase gaps between the spectra, we included data with lower SNR. This leads to a decrease in the number of mapping lines (Table\,\ref{tab:DI_info}) in the case of three images due to significant contamination by cosmics.
However, the effect of all this does not appear on the quality of the Doppler reconstructions, as we usually use a robust average of 40 mapping lines from the optical range, compared to which 30 or 35 are equally sufficient.

\begin{figure*}

\begin{tabular}{cccc}
\multicolumn{2}{c}{OR1}&\multicolumn{2}{c}{OR4}\\

\raisebox{0.73cm}{DI01}&\includegraphics[width=.41\linewidth]{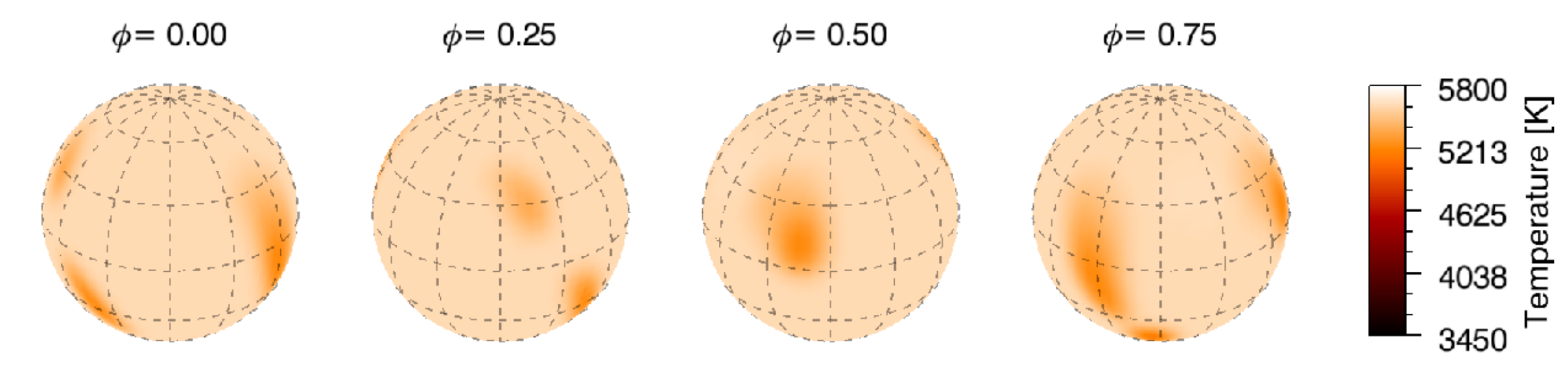} &\raisebox{0.73cm}{DI08}& \includegraphics[width=.41\linewidth]{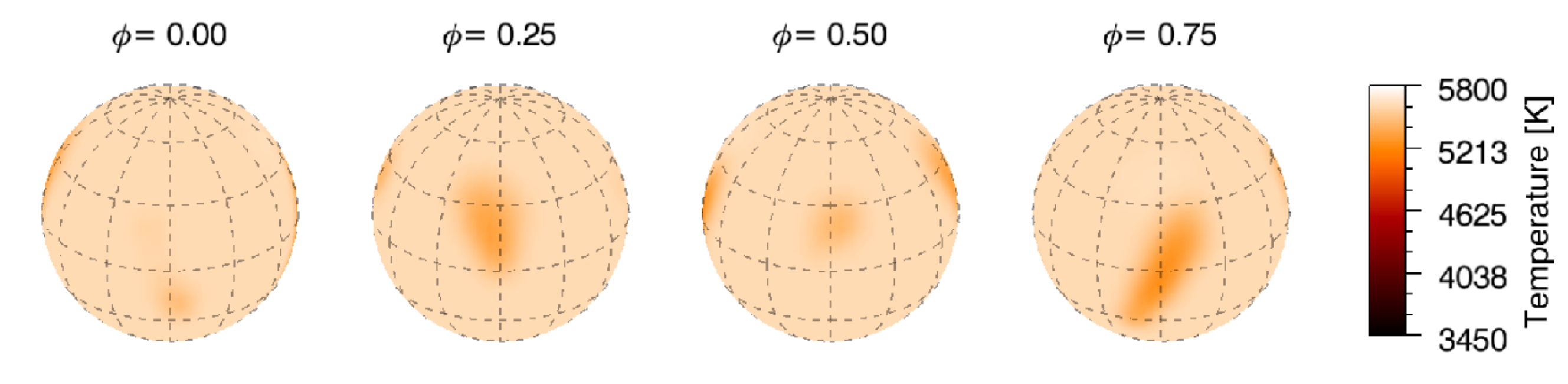}\\

\multicolumn{2}{c}{OR2}&\raisebox{0.73cm}{DI09}& \includegraphics[width=.41\linewidth]{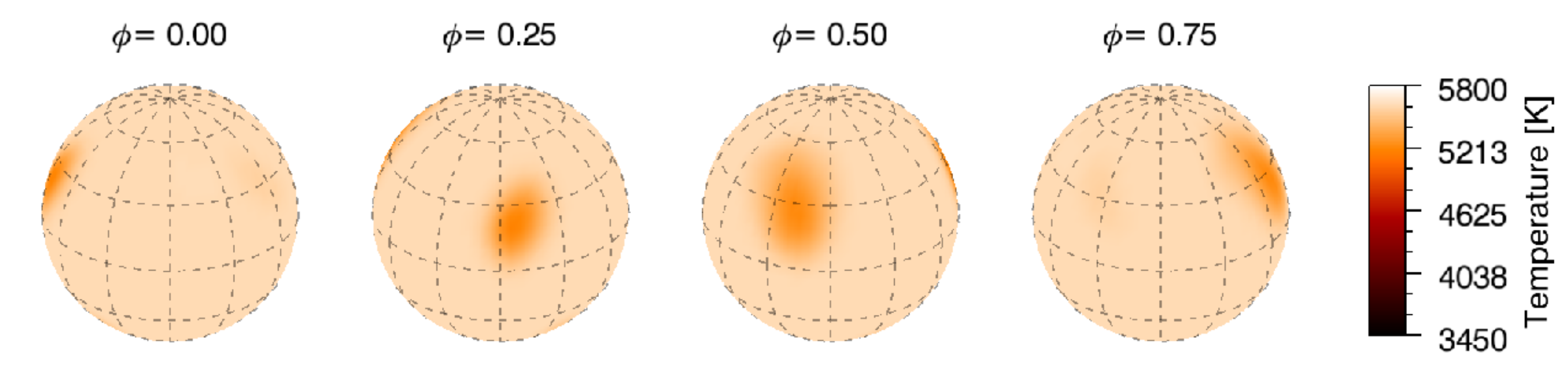} \\

\raisebox{0.73cm}{DI02}&\includegraphics[width=.41\linewidth]{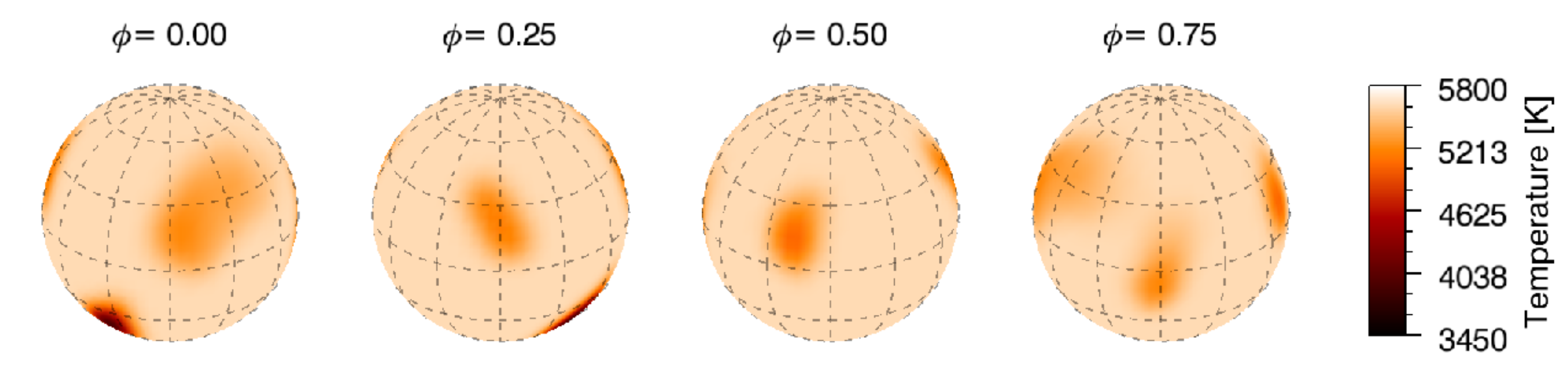} &\raisebox{0.73cm}{DI10}& \includegraphics[width=.41\linewidth]{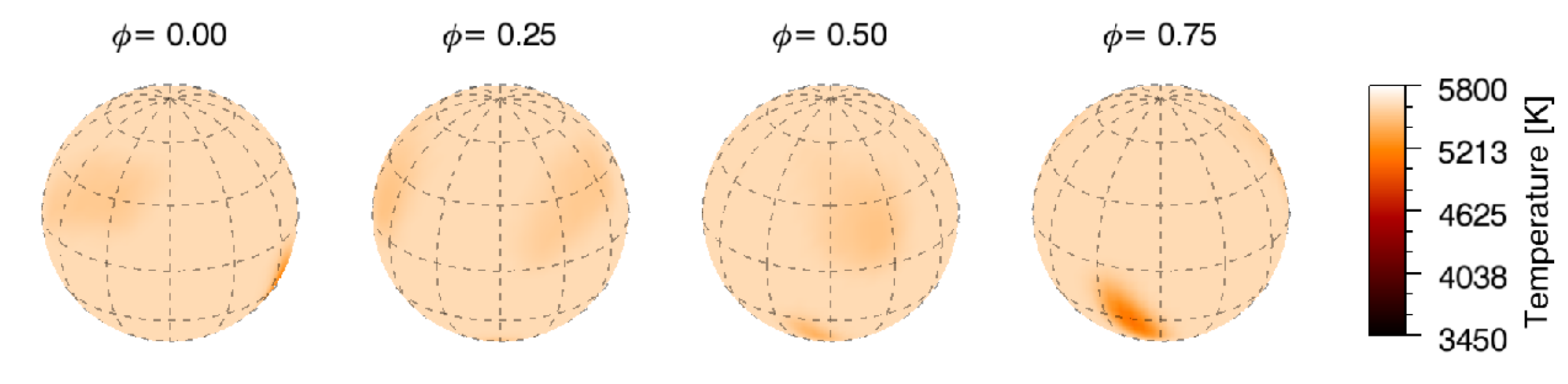}\\

\raisebox{0.73cm}{DI03}&\includegraphics[width=.41\linewidth]{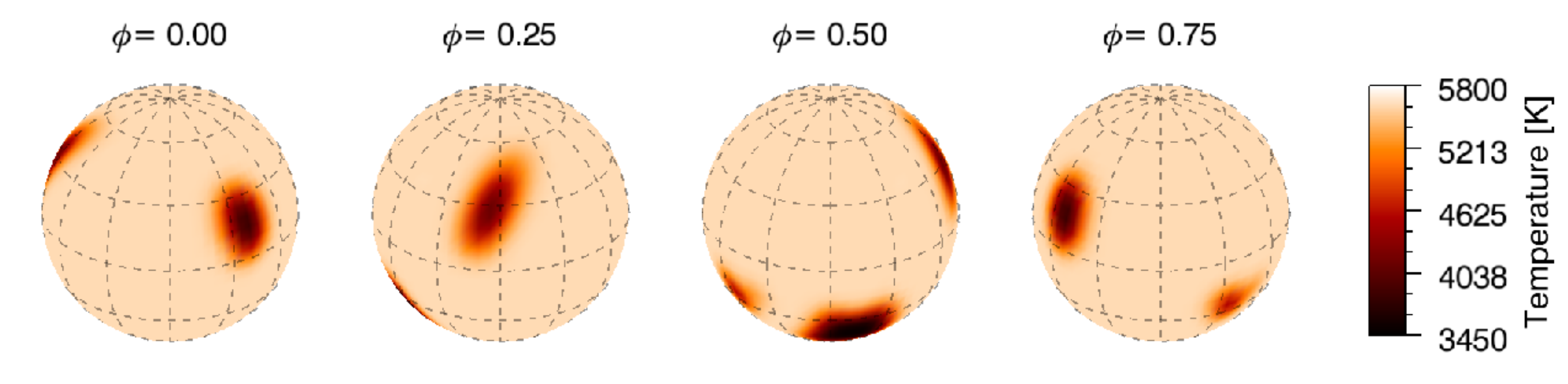} &\multicolumn{2}{c}{OR5}  \\

\raisebox{0.73cm}{DI04}&\includegraphics[width=.41\linewidth]{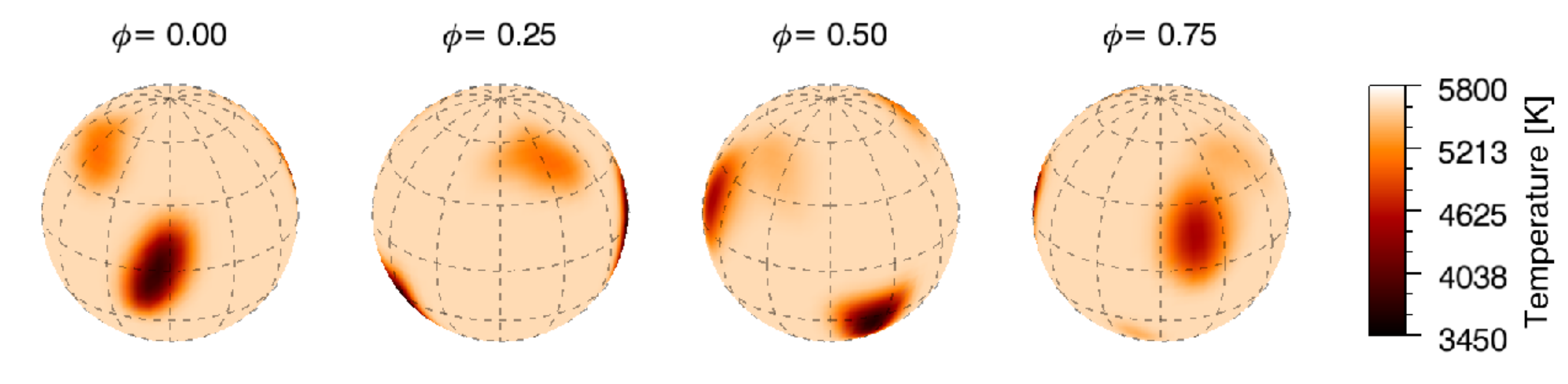} & \raisebox{0.73cm}{DI11}&\includegraphics[width=.41\linewidth]{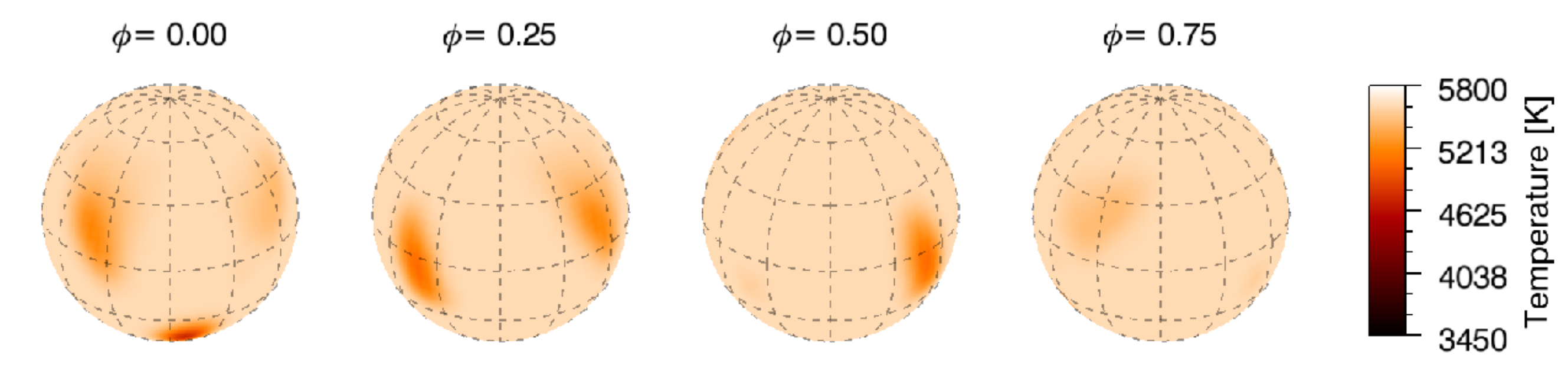}\\

\raisebox{0.73cm}{DI05}&\includegraphics[width=.41\linewidth]{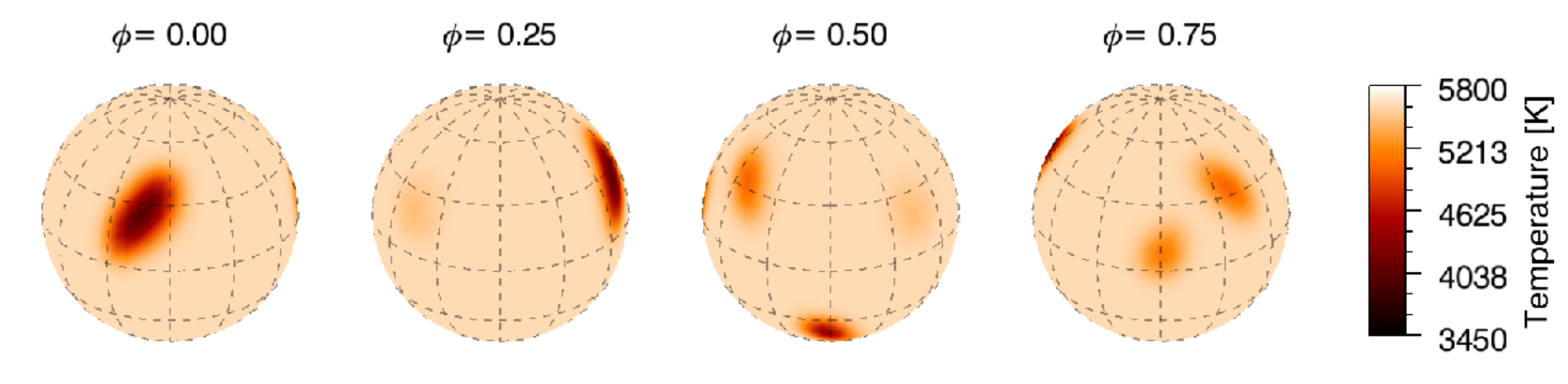} & \raisebox{0.73cm}{DI12}&\includegraphics[width=.41\linewidth]{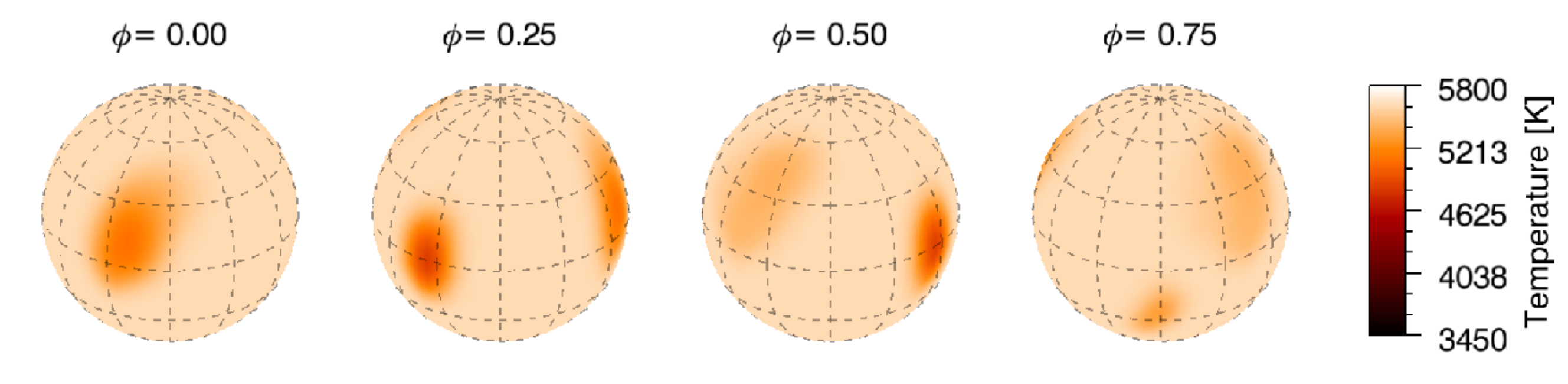}\\

\multicolumn{2}{c}{OR3}&\raisebox{0.73cm}{DI13}& \includegraphics[width=.41\linewidth]{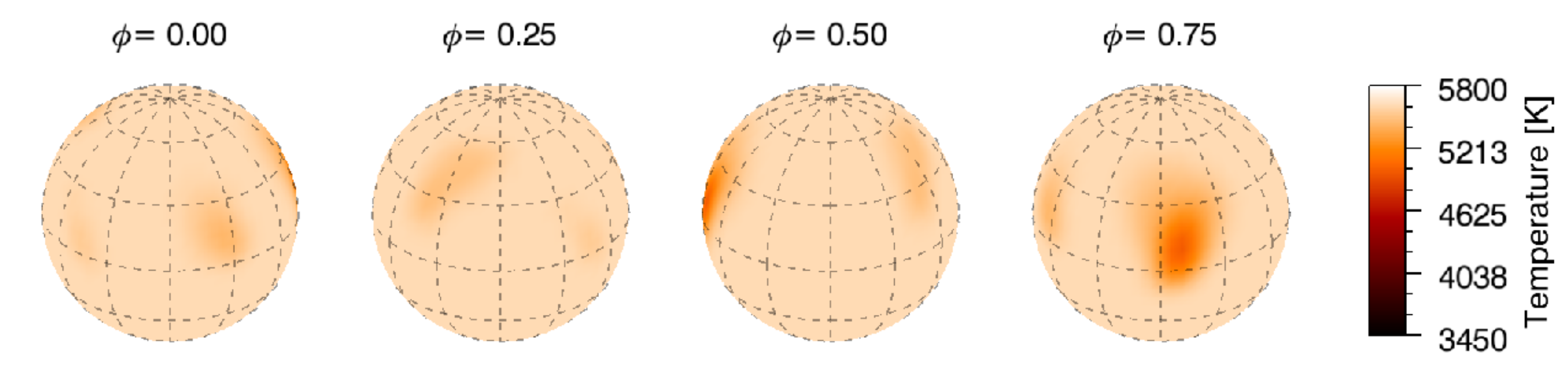}\\

\raisebox{0.73cm}{DI06}&\includegraphics[width=.41\linewidth]{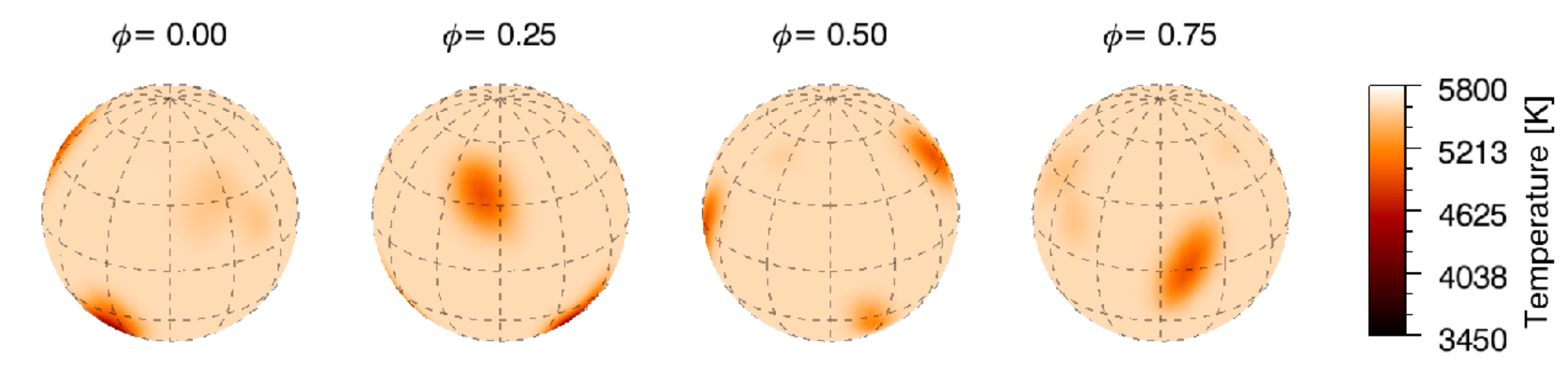} & &\\
\raisebox{0.73cm}{DI07}&\includegraphics[width=.41\linewidth]{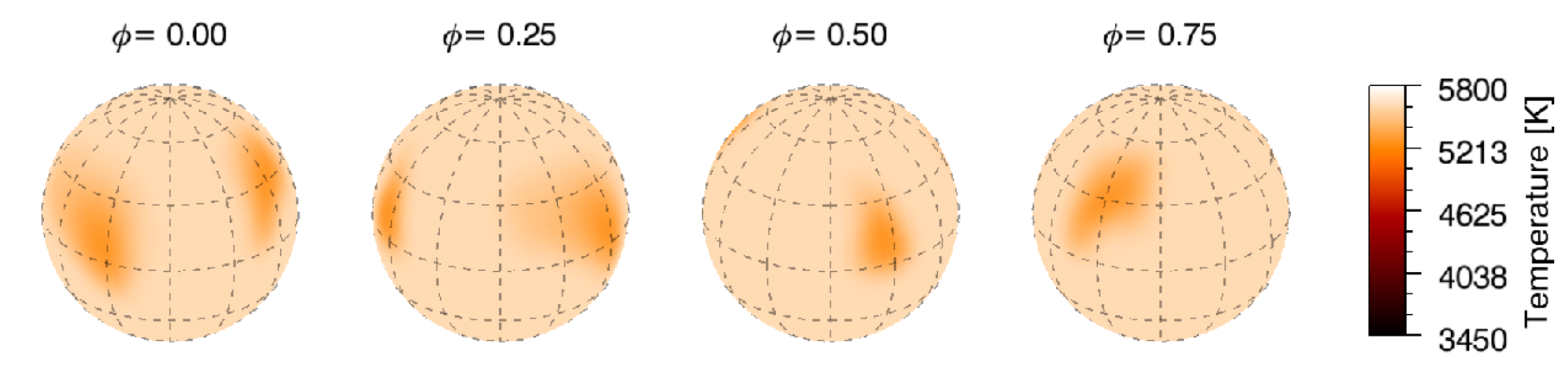} & &\\

\end{tabular}
\caption{Doppler images of EK\,Dra for the five observing runs. A total of 13 Doppler maps are displayed using the same temperature range. }
\label{fig:mosaic_dis}

\end{figure*}

To make referencing the consecutive Doppler images easier, we defined five observing runs (ORs) which include the following images: OR1 is DI01, OR2 is DI02-DI05, OR3 is DI06-DI07, OR4 is DI08-DI10, and OR5 is DI11-DI13.
Fig.\,\ref{fig:mosaic_dis} shows the resulting $13$ Doppler images and the corresponding line-profile fits are presented in Fig.\,\ref{fig:line_prof} in Appendix\,\ref{appC}. On each of the maps, multiple low-latitude spots are visible, the temperature of which is 700-1300\,K lower than that of the unspotted stellar surface. OR1 consists of one image, three spots are in phases $\phi$$\approx$$ 0.2$, $\phi$$\approx$$0.55$, and $\phi$$\approx$$0.8$. OR2 has the lowest spot temperatures and covers the most time (55 days) of all the observing runs. Even though these recovered images are not consecutive, we treat them as one observing run. A spot is consistently visible at $\phi$$\approx$$ 0.0$. Because of the time gaps between maps, it is inconclusive if these are the same spot or if decay and emergence occurred between them. At $\phi$$\approx$$ 0.25$, a spot is present on the first three images and it is the strongest on the second. The southernmost spots are at $\phi$$\approx$$ 0.5$. They appear around $-$30\degree latitude. 
Typically, Doppler maps show features on the more visible ("upper") hemisphere due to reasons inherent to the method, but since it appears on multiple maps that were reconstructed from independent data, we argue that this is a real surface feature. 

The time frame for OR3 is $12$ days and contains DI06, which is the rotation with the widest phase gap. Between DI06 and DI07 the spot configuration changes slightly. The spot at $\phi$$\approx$$0.85$ becomes more prominent and the two at $\phi$$\approx$$0.15$ and $\phi$$\approx$$0.40$ remain at about the same longitude.  There is a feature on DI06 at $\phi$$\approx$0.7, but it coincides with the phase gap and it is not visible in the following image, DI07, hence it is mostly likely an artifact.
Both OR4 and OR5 have three Doppler images covering two-two weeks of data. In the case of OR4 the spots at $\phi$$\approx$0.25 and $\phi$$\approx$0.5 become more prominent between DI08 and DI09, then weaken to DI10. The one at $\phi$$\approx$0.0 is visible in DI08 but disappears in later maps. In OR5 the spots at $\phi$$\approx$0.15 and $\phi$$\approx$0.35 diminish to DI13 while the one at $\phi$$\approx$0.7 strengthens. 

OR5 was planned to run parallel with TESS Sector 77 but due to an unexpected gap in the photometric measurements the two datasets only partially overlap. In the duration of the spectroscopic observations no flares can be observed on the light curve, so the effects of these high-energy events on the spectra can not be analyzed despite being one of the original goals of OR5. On the other hand, the spot locations on the Doppler images and those obtained from the analytic three-spot model show a good match (spots at $\phi$$\approx$0.15, $\phi$$\approx$0.35, and $\phi$$\approx$0.7).

\begin{figure}
\includegraphics[width=\linewidth]{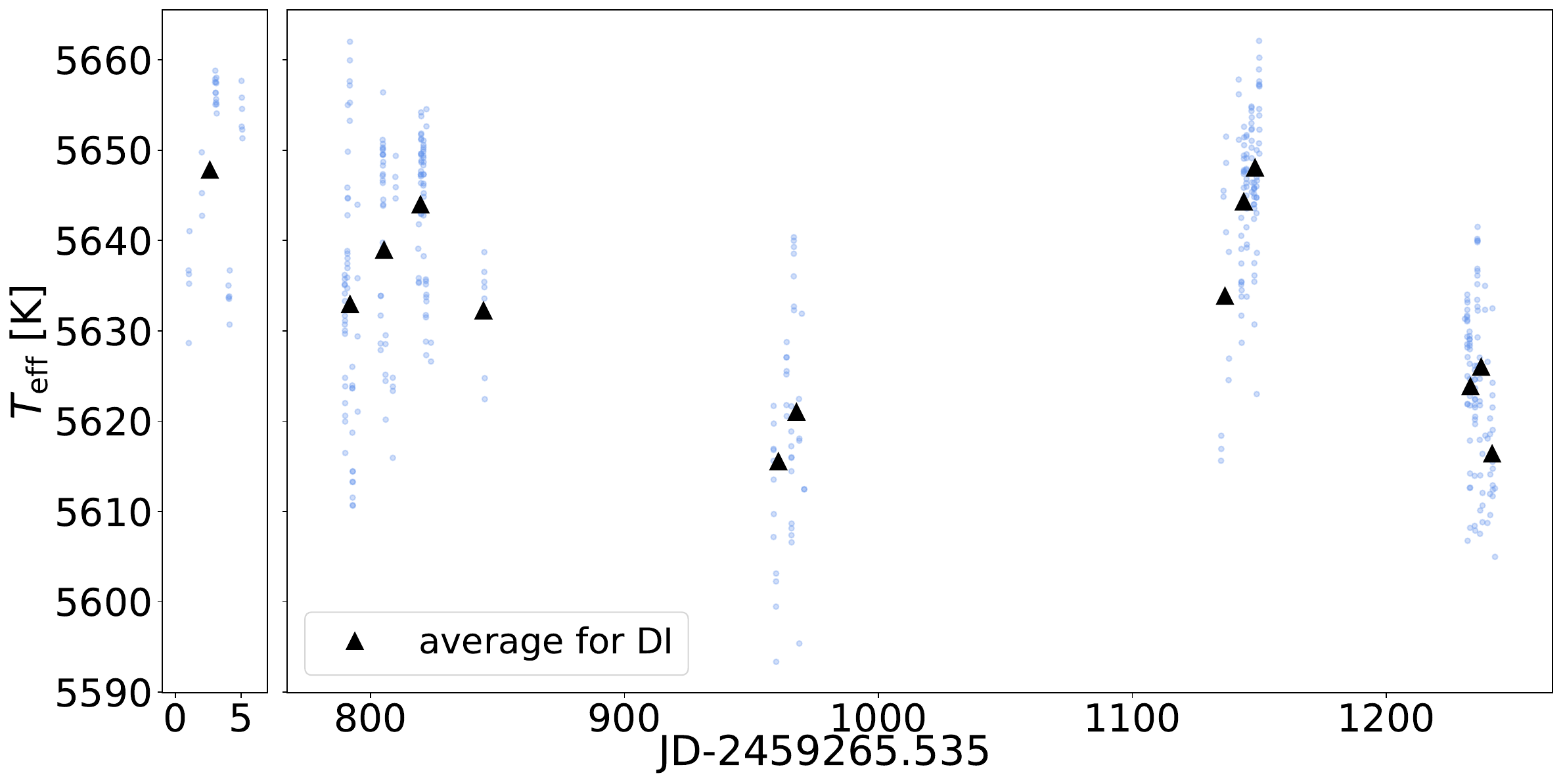}
\caption{Effective temperature from spectral synthesis for spectra with SNR higher than 80. The black triangles mark the average for each DI. The spots cause $68.7\pm 14.3$\,K variation in effective temperature. }
\label{fig:teff_periodicity}
\end{figure}

We note that the average temperature of the spots gradually decrease from OR2 to OR4. In OR5 it slightly increases again. The spot temperature change affects the average temperature of the stellar surface. This change can be quantified by SME spectral analysis. For this, spectra with signal-to-noise ratio greater than 80 were used. In the resulting Fig.\,\ref{fig:teff_periodicity} the black triangles mark the average for each DI. The spots cause a $69\pm 14$\,K variation in effective temperature. 
      
Regarding the spatial resolution of our Doppler-reconstructions, we note that with the limited spectral resolving power of $R$=21000, approximately $2\frac{R}{c}v\sin i\approx2.5$ resolution elements can be assumed across the stellar disc. Although the optimal would be at least twice this number, this is still acceptable given the circumstances \citep[cf.][]{2023A&A...674A.118S}. This is sufficiently supported by our previous tests \citep{2023A&A...674A.143K}, with which we thoroughly explored the issue of moderate resolution data; see also our further investigations in this regard in Appendix\,\ref{appA}. Finally, we mention that insufficient coverage of the rotation phase by observations (i.e. large phase gaps) can cause artifacts in the Doppler images, since the inversion lacks information about the phases that are not adequately covered by the observations \citep{2019A&A...629A.120C}. This is especially problematic when the phase gaps are larger than $\sim$0.25. Three of our images (DI04, DI05, DI06) have a phase gap of this size, all others have it below $\sim$0.2. iMap has been proven to be robust and reliable in handling occasional phase gaps \citep[e.g.][]{jarvinen2018,ei_eri}. Moreover, artifacts caused by phase gaps usually have a distinct shape, the presence of which we saw on only one of our images. Given all this, we consider the spatial resolution of our images to be moderate but acceptable, with negligible artifacts, which is also confirmed by our test results. Finally, the reliability of our images is also strengthened by the fact that Doppler maps that are within a given OR, that is, directly consecutive in time but based on completely independent data, are highly similar (apart from some traces of rapid spot evolution that occurred in the meantime).

\begin{table*}
\caption{Temporal distribution of the datasets for each individual Doppler image.}
\centering
\begin{tabular}{rcrrrc}
\hline
\hline\noalign{\smallskip}
Obs. & Image & Time interval & Time interval  & No. of & No. of \\\smallskip
Run & ID &  HJD$-$2459265.588 & [dd/mm/yyyy] & spectra & mapping lines\\
\hline\noalign{\smallskip}
OR1 & DI01  &$0$--$5.066$ & 20/02/2021--25/02/2021 & 10& 40\\
\noalign{\smallskip}
\noalign{\smallskip}
OR2 & DI02  &$ 789.777 $--$ 795.015 $ & 20/04/2023--25/04/2023 &7  &40 \\
& DI03  & $ 803.879 $--$ 808.753 $ & 04/05/2023--09/05/2023 &12 &40 \\
& DI04  & $ 809.724 $--$ 823.869 $& 19/05/2023--24/05/2023 &12 &40 \\
& DI05  &$ 841.767 $--$ 844.938 $ & 11/06/2023--14/06/2023 & 10&40 \\
\noalign{\smallskip}
\noalign{\smallskip}
OR3 & DI06  & $ 958.636 $--$ 963.841 $& 06/10/2023--11/10/2023 &10&30 \\
& DI07  & $ 965.630 $--$ 970.827 $& 13/10/2023--18/10/2023 &11 & 39\\
\noalign{\smallskip}
\noalign{\smallskip}
OR4 & DI08  & $ 1133.842 $--$ 1138.034 $ & 30/03/2024--03/04/2024 &8& 40\\
& DI09  & $ 1141.812 $--$ 1146.013 $ & 07/04/2024--11/04/2024 & 9& 40\\
& DI10  & $ 1146.819 $--$ 1149.979 $ & 12/04/2024--15/04/2024 & 8& 40\\
\noalign{\smallskip}
\noalign{\smallskip}
OR5 & DI11  & $ 1230.745 $--$ 1234.965 $ & 05/07/2024--09/07/2024 & 9& 40\\
& DI12  & $ 1235.743 $--$ 1239.864 $ & 10/07/2024--14/07/2024 & 8& 40\\
& DI13  & $ 1240.752 $--$ 1244.939 $ & 15/07/2024--19/07/2024 & 8& 35\\
\hline\\
\end{tabular}
\label{tab:DI_info}
\end{table*}

\subsection{Differential rotation}

Cross-correlation of Doppler images is a widely used method to measure surface differential rotation (\citealt{donati_ccf}). The evolution of spots (formation, dimming, interaction) can make the result of this technique unstable. To counteract this, we made use of \texttt{ACCORD} (e.g. \citealt{kovari2012}, \citealt{kovari2015}). The code uses the normalized average of the cross-correlations of consecutive Doppler images to measure differential rotation of the stellar surface by fitting the latitudinal cross-correlation peaks with a quadratic rotational law  in the form

\begin{align}
\hspace{0.3\linewidth}
\begin{array}{c}
\Omega(\beta) = \Omega_{\mathrm{eq}} (1 - \alpha_{\rm DR}\sin^2\beta),
\end{array}
\end{align}
where $\Omega(\beta)$ is the angular velocity at $\beta$ latitude, and the dimensionless surface shear parameter $\alpha_{\rm DR}=(\Omega_\mathrm{eq} - \Omega_\mathrm{pole})/\Omega_\mathrm{eq}$ is calculated from the equatorial and polar angular velocities $\Omega_{\mathrm{eq}}$ and $\Omega_{\mathrm{pole}}$, respectively. This quadratic approximation originally comes from solar physics \citep[e.g.][]{2000SoPh..191...47B}, but is widely used for spotted stars \citep[see, e.g.][and their references]{kovari_2017}.

\begin{figure}
\centering
    \includegraphics[width=\linewidth]{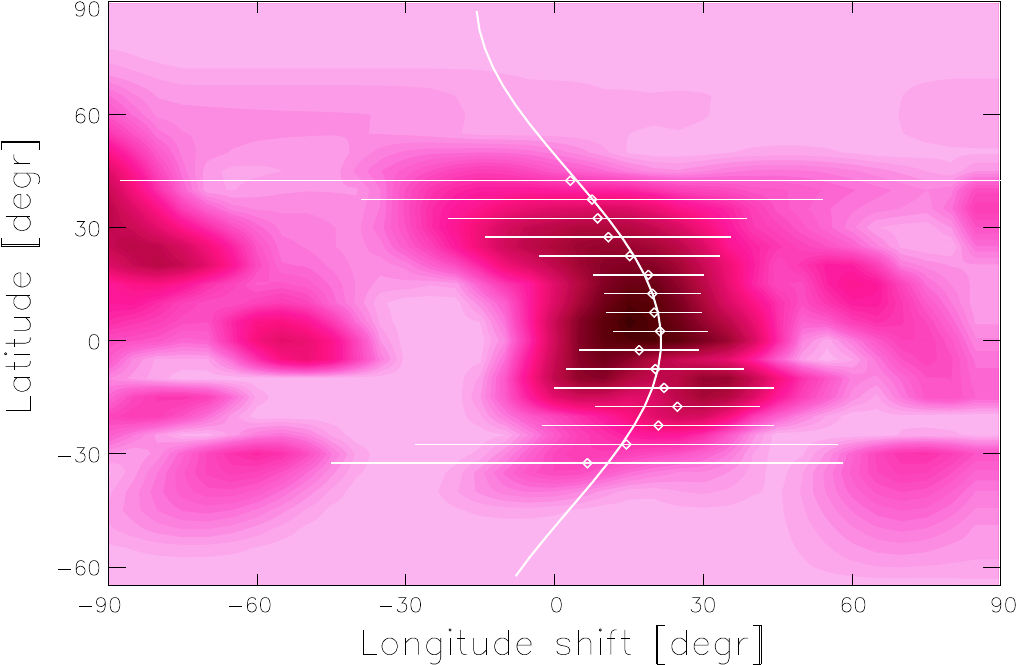}\par
\caption{Average cross-correlation function map for the consecutive Doppler reconstructions. The white circles mark the cross correlation peaks while and the white continuous line the fitted quadratic differential rotational law. The best fit surface shear corresponds to \break $\alpha_{\rm DR} = 0.030 \pm 0.008$ shear coefficient and indicates a solar-type differential rotation.}
\label{fig:accord}
\end{figure}

In OR2, OR3, OR4, and OR5 we reconstructed consecutive maps of the stellar surface. These observing runs contain three, two, three, and three time-series Doppler images, respectively, which were suitable for the cross-correlation process detailed above. The resulting differential rotation parameter is $\alpha_{\rm DR} = 0.030 \pm 0.008$ and the fit is shown in Fig.\,\ref{fig:accord}. This average differential rotation has a visible effect over 8.943 days which may also be visible on our Doppler maps.

\subsection{The variations of the H$\alpha$ line} \label{sect:Ha}

For each spectra, we measured the H$\alpha$ chromospheric activity indices ($I_{\text{H}\alpha}$) in order to examine the rotational modulation of chromospheric activity. The method for obtaining these indices was the same as in \cite{h_alpha2003}. The line index is defined in a way that shows the change in the H$\alpha$ line flux relative to the continuum flux: 

\begin{equation}
\hspace{0.35\linewidth}
I_{\text{H}\alpha}=\frac{\overline{F_{\text{H}\alpha}}}{0.5 \cdot (\overline{F_1} +\overline{F_2})}.
\end{equation}

$\overline{F_{\text{H}\alpha}}$ is the mean spectral flux in the $[-15.5,+15.5]\,\text{km}\,\text{s}^{-1}$ radial velocity interval centered on the core of the $6562.808$\,\AA\, $\text{H}\alpha$ line. $\overline{F_1}$ and  $\overline{F_2}$ are respectively the mean fluxes from the $[-700,-300 ]\,\text{km}\,\text{s}^{-1}$, and $[+600,+1000]\,\text{km}\,\text{s}^{-1}$ radial velocity intervals. 

By calculating the apparent average surface temperature of EK\,Dra over phases per thousand rotations, we recovered the temperature curve for each DI. Since the change in temperature is the consequence of stellar spots, the rotational modulation of the H$\alpha$ chromospheric activity indicator can be directly compared to the spot distribution.

\begin{figure}
\centering
    \includegraphics[width=\linewidth]{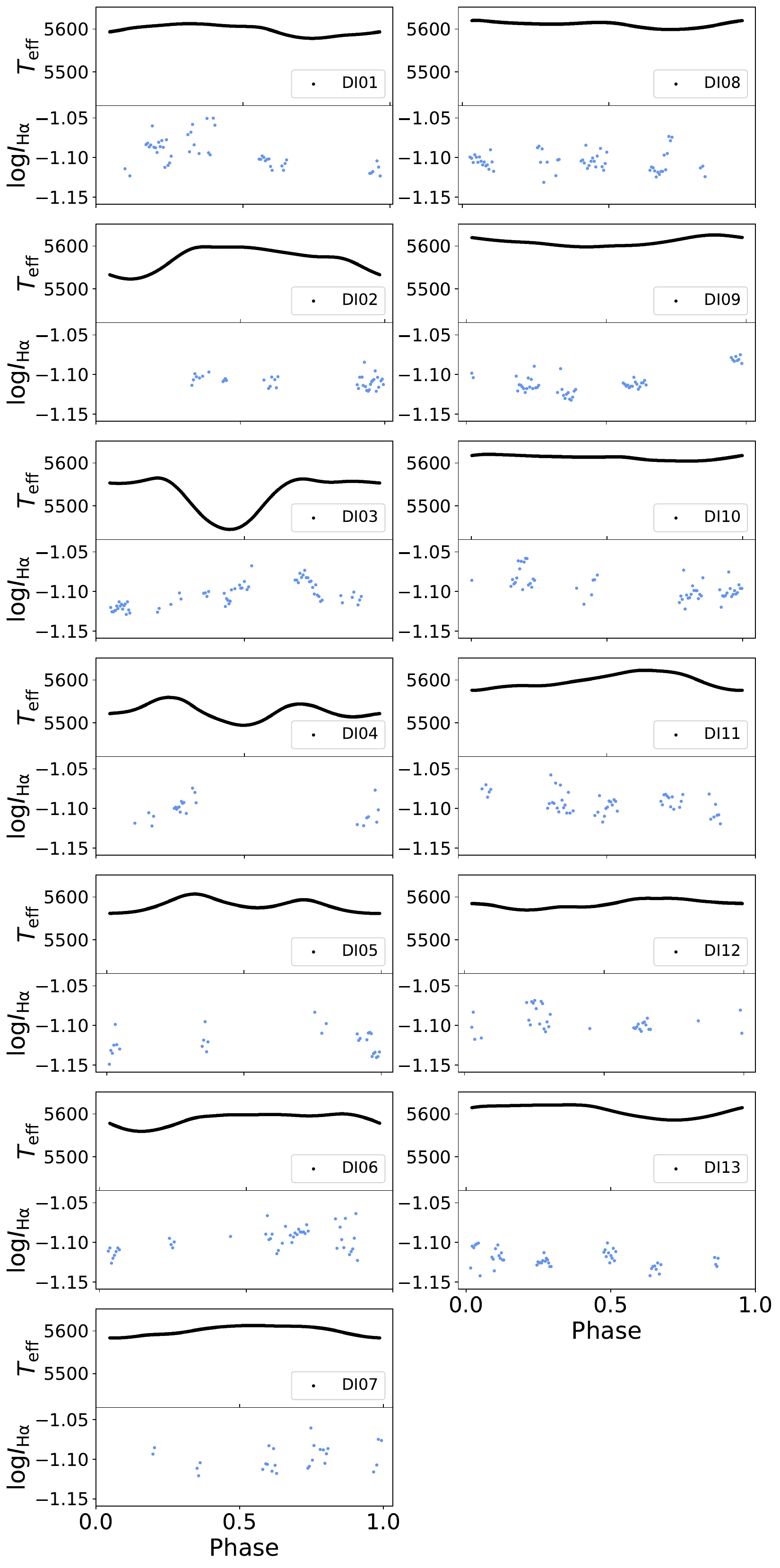}\par
\caption{The variation of the $\text{H}\alpha$ index (described in Sect.\,\ref{sect:Ha}) compared to the change in effective temperature.} 
\label{fig:Halpha}
\end{figure}

The resulting $I_{\text{H}\alpha}$ and temperature curves (Fig.\,\ref{fig:Halpha}) show changes with the rotational phase. The H$\alpha$ line is highly variable and, consequently, so is the scatter of the index. Most cases do not present a clear trend or anticorrelation with the temperature curve which we would expect based on the solar analogy. In contrast, from 2020 data of EK\,Dra \citet{Namekataetal2022a} found a clear overlap between the brightening in H$\alpha$ and the brightness decrease of the TESS light curve caused by the associated spot group. Of our Doppler images, only DI03 shows something similar: $\log I_{\text{H}\alpha}$ shows a raised value in the spot covered phases, indicating that the most chromospherically active locations are above the stellar spots.

\section{Discussion}\label{sect:disc}

\subsection{Long-term photometry and cycles}

The cyclic behavior of EK\,Dra was studied by \cite{BminV} who found a 9.2-year cycle. Later, \cite{jarvinen2018} refined this value to 8.9$\pm$0.2 years. The two studies made use of 20- and 30-year light curves, respectively (the periods covered overlapped) and also concluded that the star had continued to fade since 1985 until the end of the observations.

We studied the long-term photometric behaviour of EK\,Dra based on the Sect.\,\ref{section:phot_obs} light curve spanning over 120 years. The star shows cyclic behaviour with a period of 10.7-12.1 years, which reinforces the findings of the above-mentioned studies and shows that this cycle has been persistent for the last century. We found an additional cycle of 7.3-8.2 years period that appeared around 1950-1960. This is also the point where the star started its dimming, which has been going on for about $\sim$64 years. Considering the preceding 60-year brightening phase, one possible interpretation of the data suggests a cycle longer than a century. However, such long-term brightening do not necessarily imply cyclic variation, as pointed out by \citet{2014A&A...572A..94O}.
A possible explanation for a magnetically-driven, non-cyclic, but trend-like, long-term brightness change of the red giant XX\,Tri was recently suggested by \citet{2024NatCo..15.9986S}. According to them the continuously blocked flux by large cool starspots can be redistributed on a global scale rather than locally through an increase of, for example, faculae activity. It thereby alters the overall brightness of the star on a time scale likely much longer than the stellar rotation, even over decades.

While the binary component could also be the cause of the long-term brightness variation, we argue that this is not the case here. Based on \cite{2005A&A...435..215K}, a periastron was around 1987 and the previous one between 1937-1947. The earlier of these occurred when the star was brightening, the other when it was dimming. These dates are well covered with data points, so the different trends are clearly visible, and at the two periastron no similar change can be seen in the light curve. The STFT analysis cannot detect the periodicity caused by the binary component, since the possible STFT peak by the binary period overlaps with the peak caused by the gap in the data -- both result in a peat at $\sim$50 years period -- making the two undistinguishable.
     
\begin{figure}
\centering
    \includegraphics[width=\linewidth]{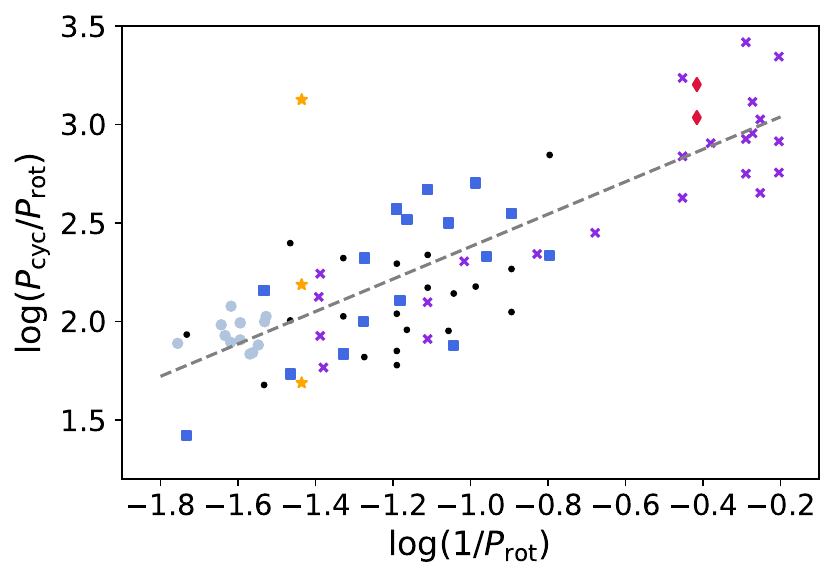}\par
\caption{The relation between rotational and cycle periods. The cycles we obtained from the long-term photometry data for EK\,Dra are marked by red diamonds. The cycles for stars denoted by purple are from \citet{2009olah}. The rest are adopted from \citet{magnetic_cycles} where the authors differentiated between simple cycles (light blue circles), complex cycles (dark blue squares) and additional cycles (black dots) to the complex ones. The solar cycles (Gleissberg, Schwabe, and 3–4-yr cycles) are marked in orange.}
\label{fig:rot_cyc}
\end{figure}

In order to place the activity cycles of EK\,Dra in the context of other active stars, in Fig.\,\ref{fig:rot_cyc} we took the $P_{\text{rot}}$ rotational and $P_{\rm cyc}$ cycle periods from \cite{2009olah} and \cite{magnetic_cycles}, plotted the relation between $P_{\rm cyc}/P_{\rm rot} $ and $1/P_{\rm rot}$ and included EK\,Dra in the sample. The cycles derived for EK\,Dra are consistent with that of fast rotators.

\begin{figure}
\centering
    \includegraphics[width=\linewidth]{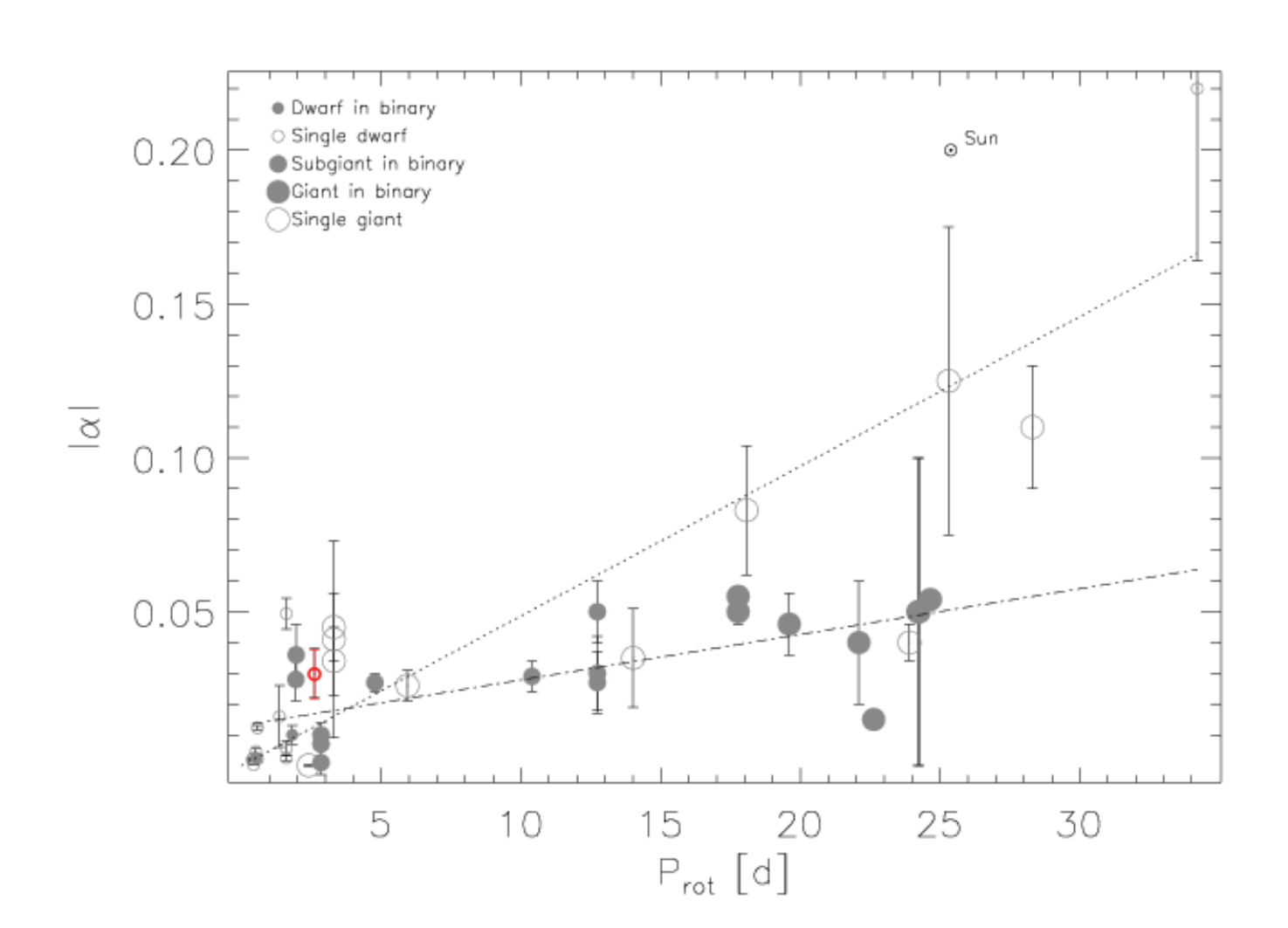}\par
\caption{The absolute value of the surface shear coefficient $\alpha_{\rm{DR}}$ as a function of the rotational period for spotted stars \citep{kovari_2017}. Single stars are indicated by circles, while those in binary systems are denoted by filled symbols. EK\,Dra is marked in red. The dotted line is a linear fit to single stars with a slope of $|\alpha_{\rm DR}| \propto 0.0049\,P_{\rm rot}$[d], the lower slope of the dash-dotted line is $|\alpha_{\rm DR}| \propto 0.0014\,P_{\rm rot}$[d] and it is fitted to stars in binary systems \citep{2023A&A...674A.143K}.} 
\label{fig:a_vs_prot}
\end{figure}

\subsection{Flare activity}

The flare energy distribution of EK\,Dra can be described by a broken power-law. The Sun \citep{2003A&AT...22..325K} and some dwarf stars \citep{2018ApJ...858...55P} also exhibit the same behaviour. This break in the power-law might be caused by statistical fluctuations. At the same time the breaking point was interpreted by \cite{2018ApJ...854...14M} as a critical energy above which the size of the magnetic loop, which becomes twisted and releases its energy in the shape of flares, becomes higher than the local scale height. This critical energy is changing from star to star and for EK\,Dra it is around $E=10^{34}$\,erg. We derived a power-law index of $ 1.466 \pm 0.007$ below this energy. \cite{Namekataetal2022a} conducted a similar analysis for TESS Sectors 14-16 and 21-23. The slope of our FFD between $E = 10^{33}$ and $E = 10^{34}$ matches theirs within the errors.

In Sect.\,\ref{sect:flares} we investigated the longitudinal flare distribution and compared the distribution of flares to that of the spots by applying the Kuiper test and the two-sample Kolmogorov$-$Smirnov test, respectively. Neither of the cases can reject the null hypothesis of a uniform flare distribution.

\subsection{Spot activity and differential rotation}

From rotation to rotation, the spot configuration on the surface of EK\,Dra changes. This variation -- modelled with an analytic three-spot model (Figs.\,\ref{fig:sector14}, \ref{fig:S15-22}, \ref{fig:S23-49}, \ref{fig:S75-77}) -- appears as both decay and emergence of spots, as well as their shift in position. The deviation of the analytic model from the light curve can be attributed to either smaller spots or the spots being approximated with perfect circles during modeling. 
In addition, the longitudinal positions of the spots, which can be derived relatively precisely from photometric spot models, show trend-like changes, suggesting that the star performs surface differential rotation.

Surface changes of EK\,Dra on the rotational time scale can also be confirmed with the Doppler images (Fig.\,\ref{fig:mosaic_dis}), especially those that are immediately consecutive in time. The reconstructed spots appear mostly at mid-latitudes, around 30-60$^\circ$, but they also show up at low latitudes, and a few cross the equator (see, e.g. DI12). The numerical simulation of flux emergence \citep[][]{schussler1996,granzer2000,isik2018} could not reproduce these low-latitude spots, thus predicting a "zone of avoidance" between $\pm$20$^\circ$. \citet{senavci2021} found that information on activity in the less visible hemisphere at mid-latitudes can leak and cause these low-latitude spots. Unlike these, the flux emergence simulation in \citet{low_lat_spots}  reinforces these spots as real surface features on fast rotating solar-type stars. The variation of the surface also appears in the effective surface temperature obtained by SME (Fig.\,\ref{fig:teff_periodicity}). The spots cause a 69$\pm$14\,K variation in effective temperature.
We note that the relatively low resolution of our instrument has an effect on the result of the Doppler inversion. This is tested and discussed in Appendix \ref{appA}.
   
EK\,Dra has a long history of Doppler imaging. The first such study was carried out by \citet{strassmeier_rice_1998} and ever since, multiple Doppler images were obtained \citep[e.g.][]{jarvinen2007,2017MNRAS.465.2076W,jarvinen2018,namekata2024}. These studies analyze one to four maps, while our campaign reconstructs $13$ images with multiple consecutive ones. This provides an opportunity to follow the short-term changes on the surface and to recover the differential rotation shear parameter $\alpha_{\rm DR}=0.030 \pm 0.008$. Figure\,\ref{fig:a_vs_prot} shows that this value obtained for EK\,Dra fits to the trend observed for stars with different rotational periods. We note that earlier attempts were made to recover the differential rotation parameter, with resulting values being an order of magnitude lower ($\alpha_{\rm DR}=0.00091\pm 0.00001$, \citealt{jarvinen2007}) or higher ($\alpha_{\rm DR} = 0.108 \pm 0.0554$, \citealt{2017MNRAS.465.2076W}) than ours.

Linear and non-linear spot decay models (e.g. \citealt{1997SoPh..176..249P,2021ApJ...908..133M}) originate from the observational fact that the lifetimes of sunspots depend on their area. When considering decay models, the spots on EK\,Dra can be considered either as monolithic spots or clusters of smaller spots, since the resolution of our Doppler imaging does not reach the typical size of sunspots. Based on the Doppler maps, the lower limit for the lifetime of stellar spots is around 10-15 days. \cite{v815her} reported a similar value in the V815\,Her system for the Aa component, which is a $30$\,Myr solar analogue. \cite{2007A&A...464.1049I} found that while large monolithic spots have a lifetime of months, a cluster of small ones decays on a time scale of a few days to a few tens of days. Although the lifetime of the spots on EK\,Dra matches that of the spot clusters better, \citet{jarvinen2018} found no evidence of such conglomerate. \citet{2019ApJ...871..187N} report that on rapidly rotating stars the spots evolve on a shorter timescale, so the rotational period of EK\,Dra is possibly one of the key factors in its spots lifetime.

\section{Conclusions}

  In this study, we used photometric and spectroscopic data from the young solar analogue EK\,Dra to investigate the magnetic activity of the star on shorter and longer time scales. The summary of our main conclusions is the following.
  \begin{itemize}
    
    \item From long-term photometric data we find that EK\,Dra exhibits a $10.7-12.1$ year cycle that was persistently present for 120 years. In addition, the star shows a 60 years-long dimming phase preceded by a brightening one, which points to a possible cycle period longer than $120$ years.
    
   \item We modelled light curves from 13 TESS sectors using an analytical three-spot model. From the longitudinal shifts of the spots we infer surface differential rotation.
    
    \item We found a total of $142$ flares in the TESS data. Their longitudinal distribution shows no correlation with the rotational phase or with the spotted longitudes. The flare energy distribution follows a broken power law with indices $1.466\pm0.0007$ for the $E=10^{33}-10^{34}$\,erg range and $2.335\pm0.11$ above $ E=10^{34}$\,erg.
    
    \item The $13$ reconstructed Doppler images show that the spotted surface of EK\,Dra varies from rotation to rotation, with the lifetime of the spots being 10-15 days.
    
    \item Consistent with the presence of differential rotation suggested by photometric spot modeling, Doppler imaging shows that the star exhibits solar-type differential rotation with a surface shear parameter of $\alpha_{\rm DR}=0.030\pm 0.008$. This value, obtained by cross-correlation of consecutive Doppler images, is in good agreement with the $\Delta P/\overline P_{\mathrm{rot}}=0.02 \pm 0.02$ value, which is a rough estimate of the shear parameter from the rotation period changes.
    
    \item The H$\alpha$ line was shown to be highly variable, so the derived activity indices have a large scatter. This makes the connection between photospheric and chromospheric activity inconclusive.  
  
  \end{itemize}

Further photometric observations could confirm the long-term brightening and dimming as part of an activity cycle or as a change related to the evolution of EK\,Dra. Moreover, the relationship between the photosphere and the chromosphere is also a subject of further research, since, as we have now seen in the case of EK\,Dra, chromospheric tracers are not always closely associated with the photospheric spots.

\begin{acknowledgements}

The authors thank the anonymous referee for improving the quality of the paper with helpful comments and suggestions. This research was funded by the Hungarian National Research, Development and Innovation Office grant \'Elvonal KKP-143986. 
Authors acknowledge the financial support of the Austrian-Hungarian Action Foundation grants 98\"ou5, 101\"ou13, {104\"ou2,} 112\"ou1.
L.K. acknowledges the support of the Hungarian National Research, Development and Innovation Office grant PD-134784. 
K.V.  is supported by the Bolyai J\'anos Research Scholarship
of the Hungarian Academy of Sciences.
On behalf of the \textit{"Looking for stellar CMEs on different wavelengths"} project we are grateful for the possibility of using HUN-REN Cloud \cite{MTACloud} which helped us achieve the results published in this paper.

This work has used data provided by Digital Access to a Sky Century @ Harvard (DASCH), which has been partially supported by NSF grants AST-0407380, AST-0909073, and AST-1313370.

This paper includes data collected by the TESS mission. Funding for the TESS mission is provided by the NASA's Science Mission Directorate.

\end{acknowledgements}

\bibliographystyle{aa}
\bibliography{bib}

\appendix

\section{Comparison with PEPSI data and the effect of spectral resolution}\label{appA}

\begin{figure}
\includegraphics[width=\columnwidth]{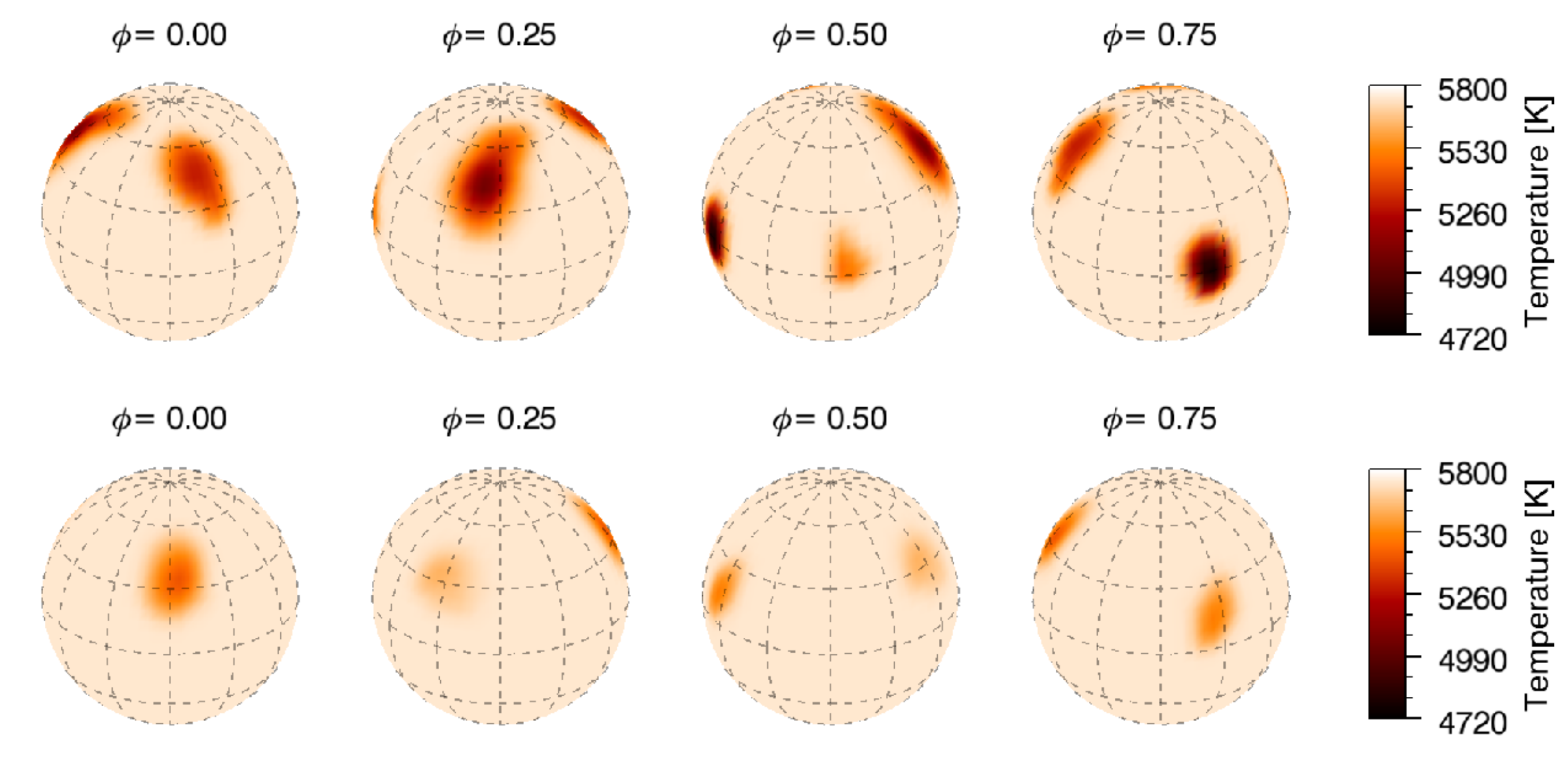}
\caption{Doppler images of EK\,Dra using the dataset published in \cite{jarvinen2018}. The upper panel shows a map with the resolution and inversion parameters used in \cite{jarvinen2018}. The lower panel contains the map with the resolution downgraded to 20\,000 and done with the inversion parameters from this paper.}
\label{fig:pepsi}
\end{figure}

\begin{sidewaysfigure*}
\begin{tabular}{cc}
\includegraphics[width = 0.48\linewidth]{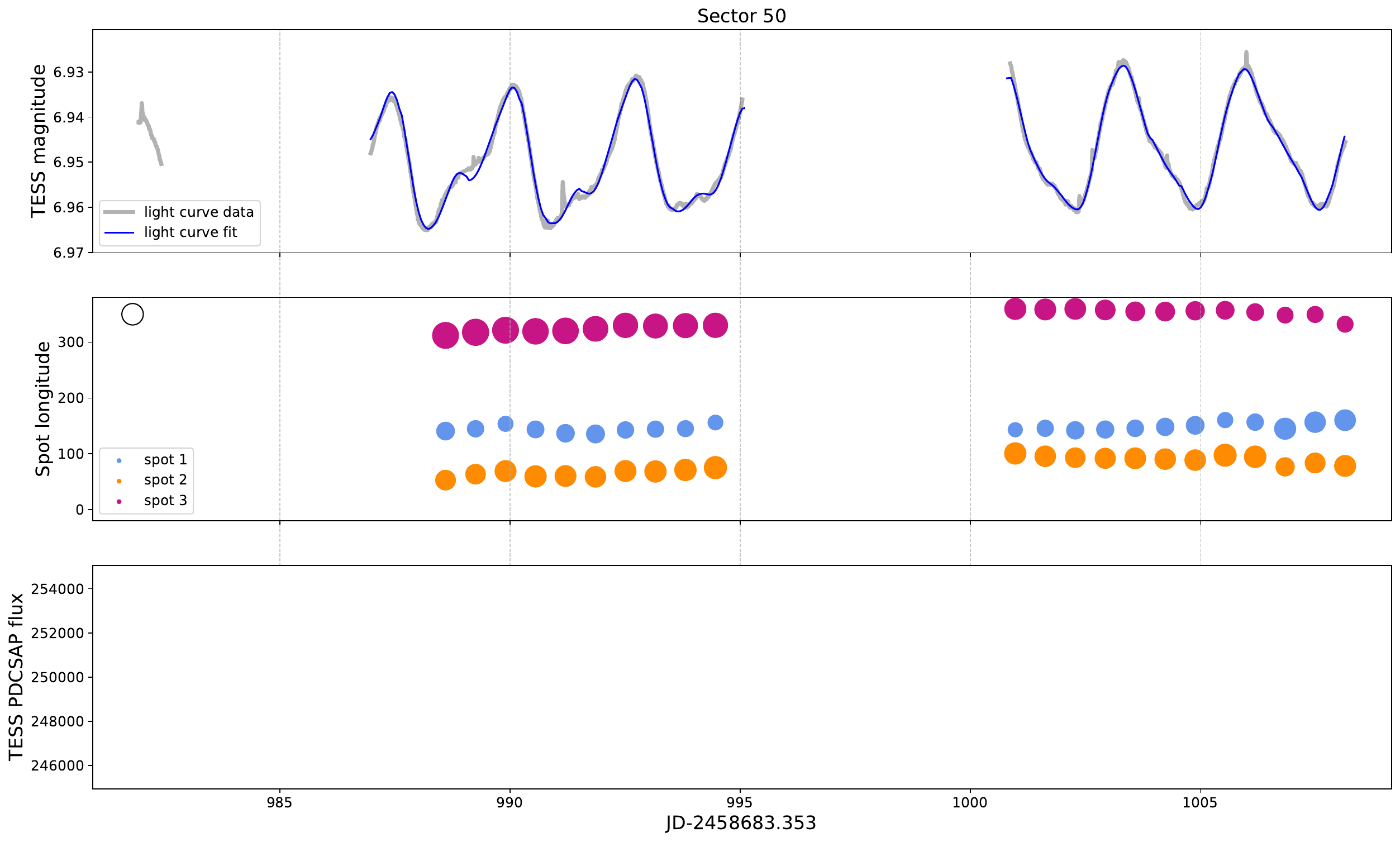} &
\includegraphics[width = 0.48\linewidth]{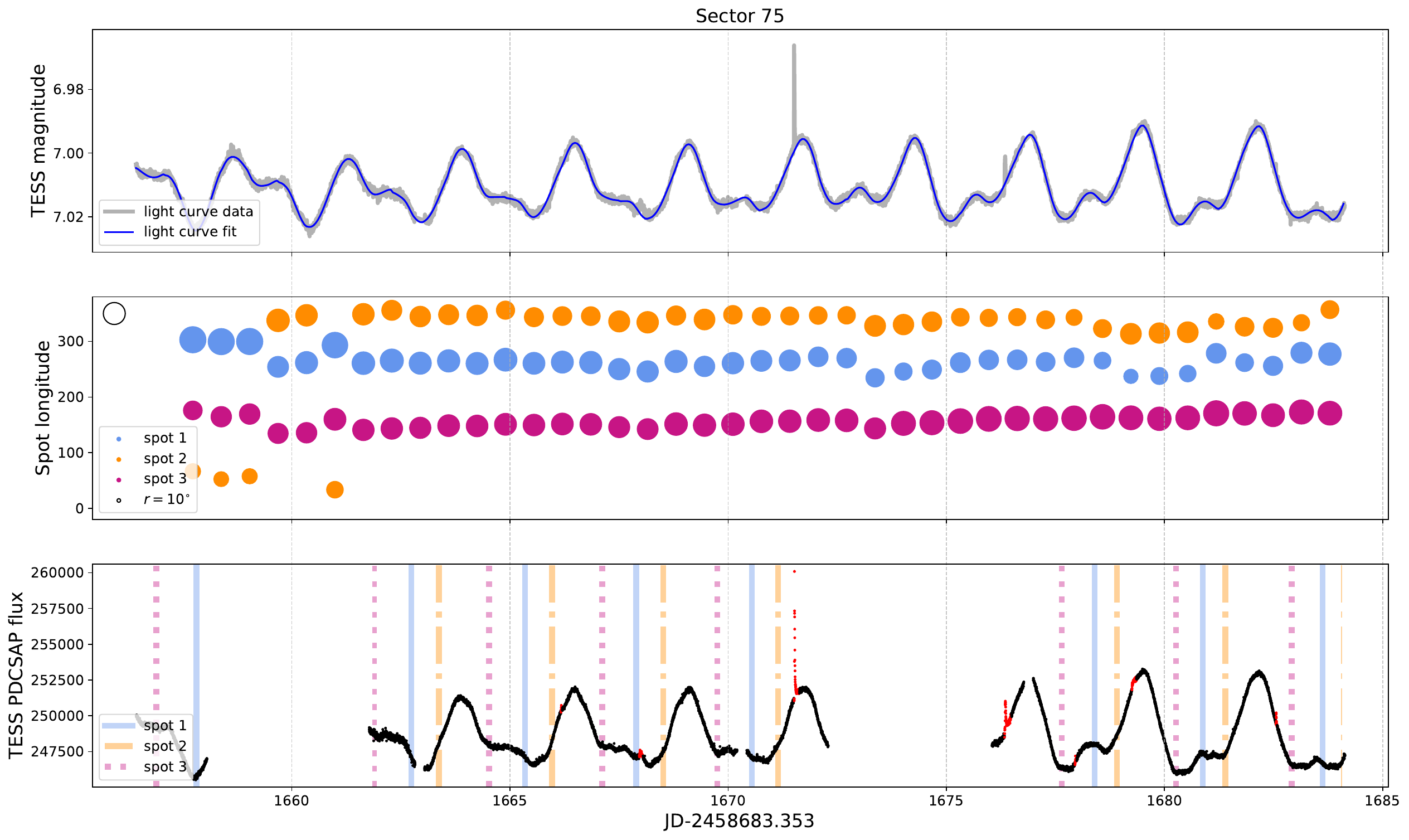} \\
\includegraphics[width = 0.48\linewidth]{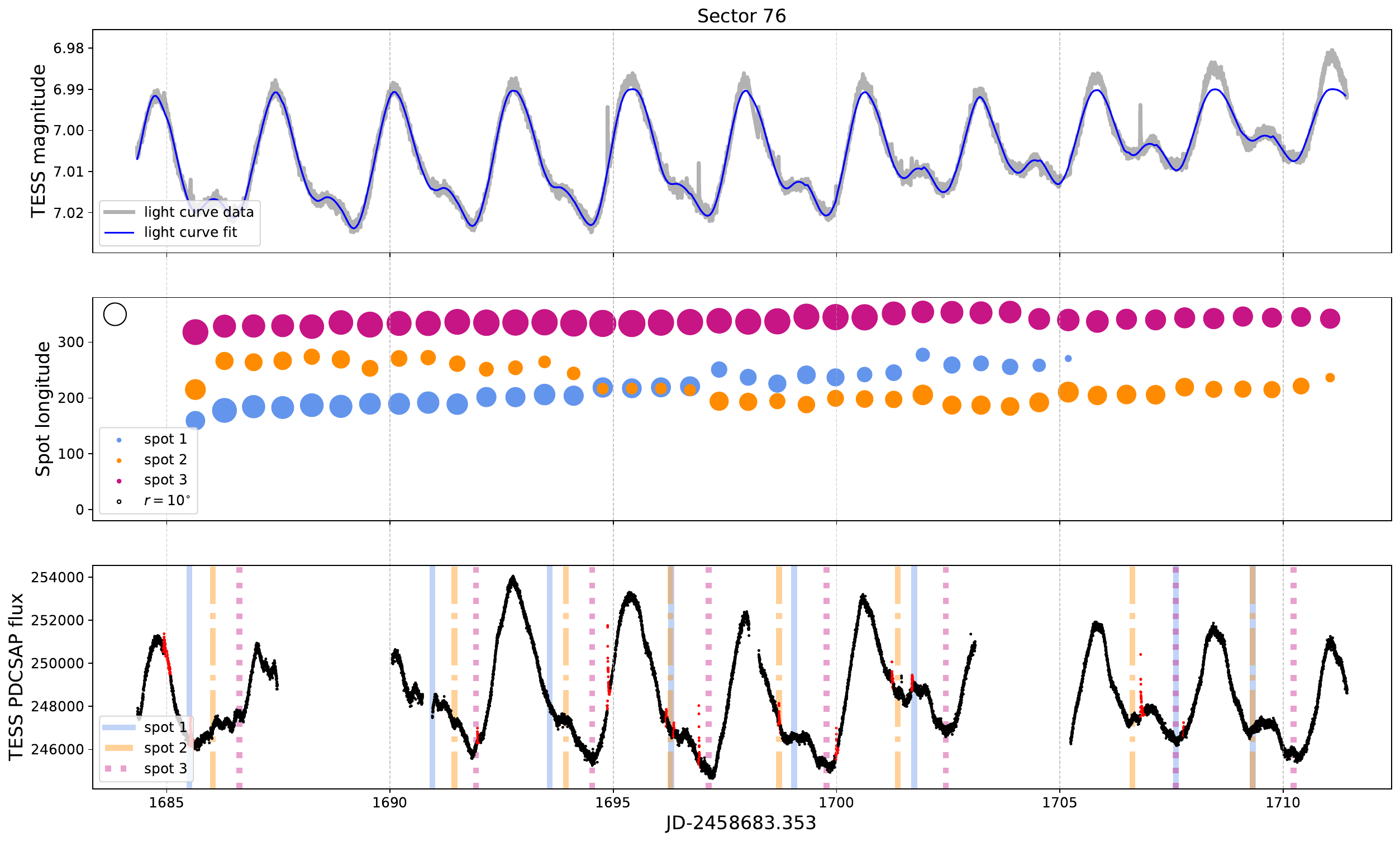} &
\includegraphics[width = 0.48\linewidth]{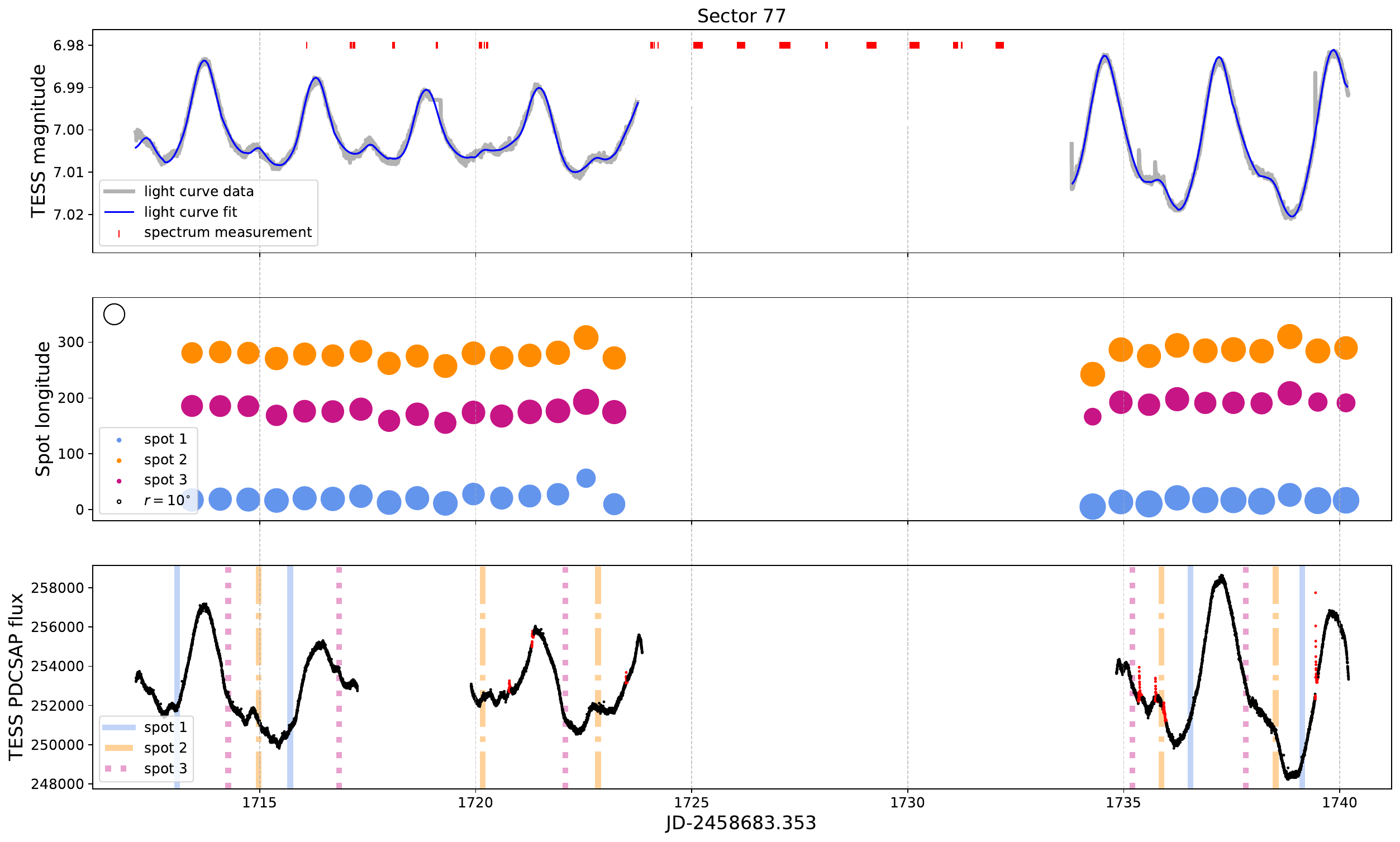} 

\end{tabular}
\caption{Same as Fig.\,\ref{fig:sector14} but for TESS sectors S75, S76, S77. In the plot corresponding to Sector 77, the top panel shows the time of the spectrum measurements marked with red. For Sector 50 no TESS PDCSAP flux is available.}
\label{fig:S75-77}
\end{sidewaysfigure*}

In addition to the Piszk\'estet\H{o} observations, the spectra published in \citet{jarvinen2018} were used for validation purposes. Using the method detailed in \citet{2023A&A...674A.143K} the $R=230\,000\pm30\,000$ resolution PEPSI spectra were downgraded to $R=20\,000$ before Doppler imaging. In the resulting map (Fig.\,\ref{fig:pepsi}), the location of the spots matches that of \citet{jarvinen2018}. However, it is apparent that small-scale structures are suppressed by the lower resolution; the smallest spot reconstructed from the original PEPSI data at $\phi$$\approx$$0.5$, with an area of 3\% and a temperature difference of $\Delta T=280\,\mathrm{K}$ cannot be recovered after the resolution degradation. The shape of larger structures is somewhat simplified and the temperature contrast decreases with decreasing resolution, but their position is preserved and Doppler reconstructions are still suitable for studying short- and long-term spot evolution. Moreover, by averaging cross-correlation function maps of consecutive pairs of Doppler images, surface differential rotation can also be detected, as demonstrated by the tests of \citet{2023A&A...674A.143K}. We note that despite the lower resolution, short-term spot evolution and the gradual change of spot temperatures are clearly visible on the subsequent maps in Fig.\,\ref{fig:mosaic_dis}. Finally we also mention that the reliability of our Doppler reconstructions is supported by the agreement between our DI08 image and the analytical three-spot model obtained for the temporally coincident TESS S77 data (cf. Figs.\,\ref{fig:mosaic_dis} and \ref{fig:S75-77}).

\section{Analytic spot model for all available TESS sectors}\label{appB}

\begin{sidewaysfigure*}
\begin{tabular}{cc}
\includegraphics[width = 0.48\linewidth]{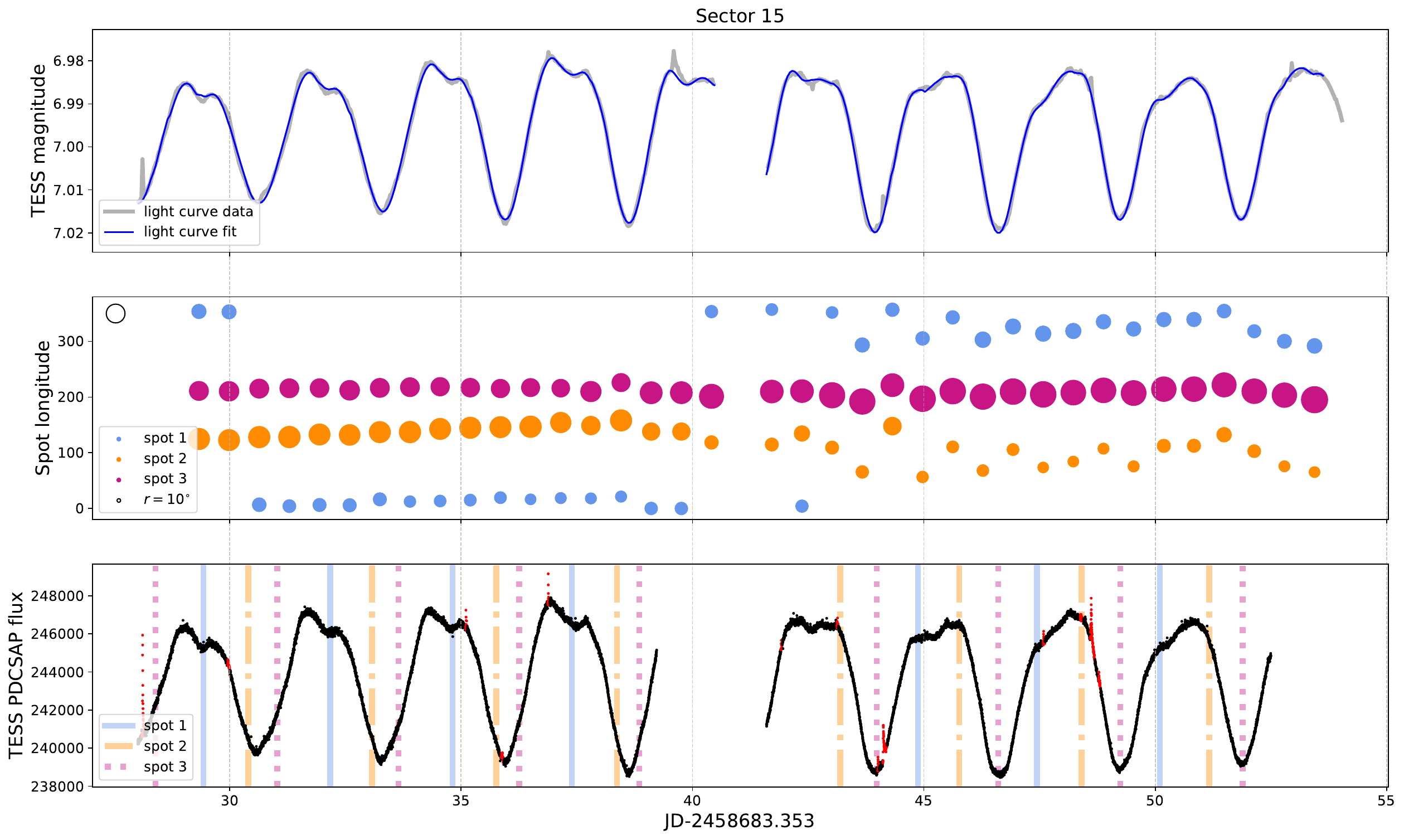} &
\includegraphics[width = 0.48\linewidth]{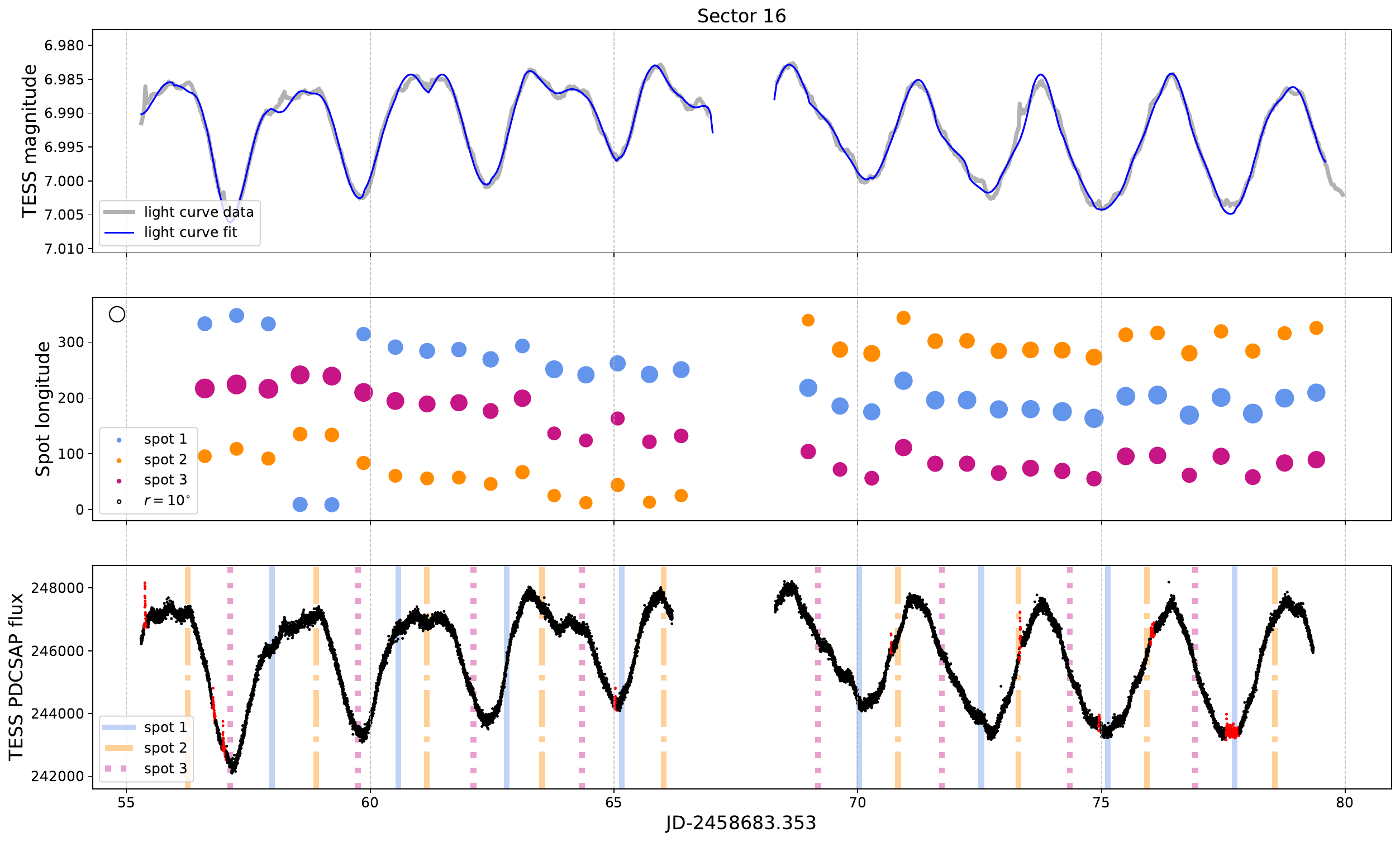}\\
\includegraphics[width = 0.48\linewidth]{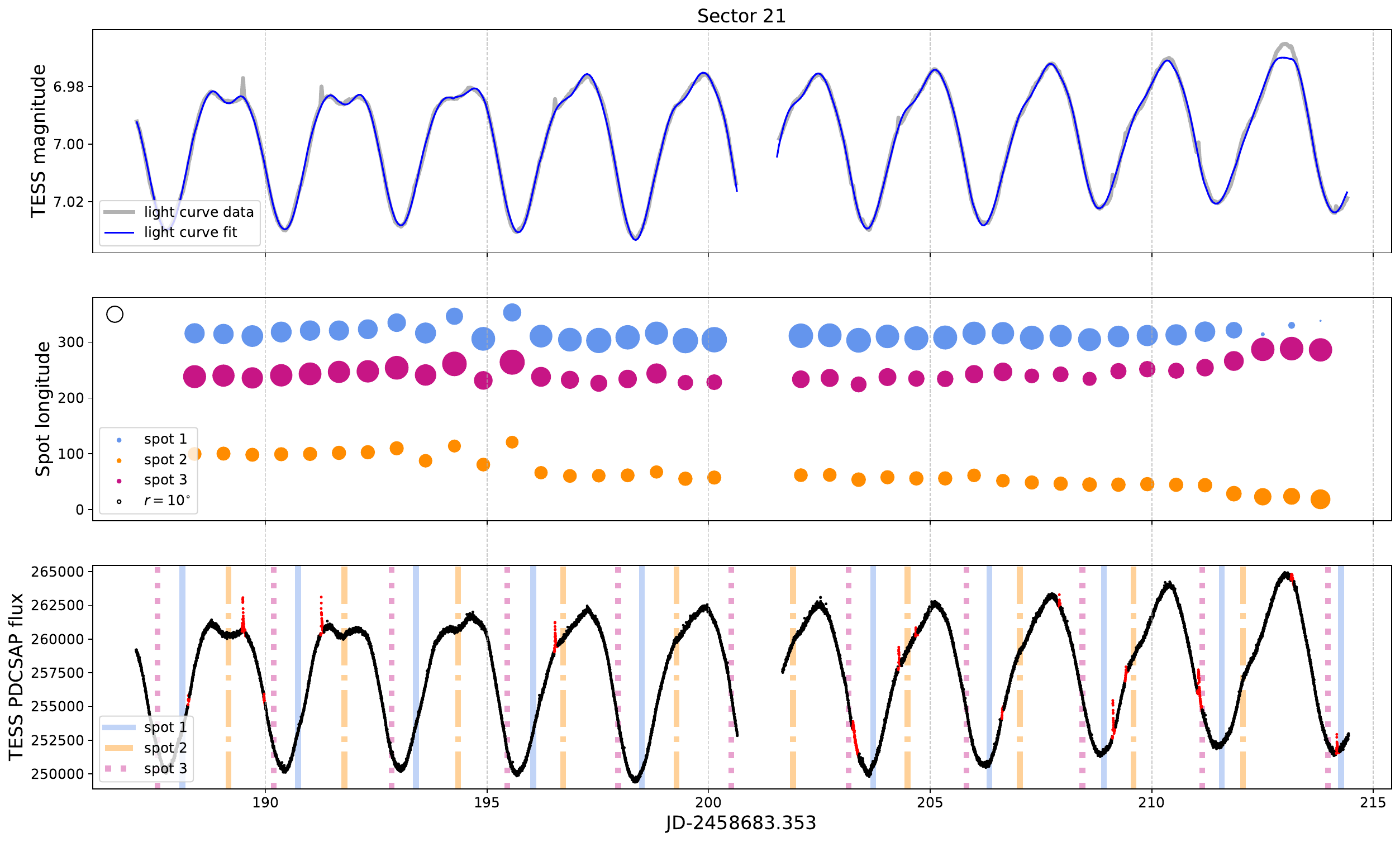} &
\includegraphics[width = 0.48\linewidth]{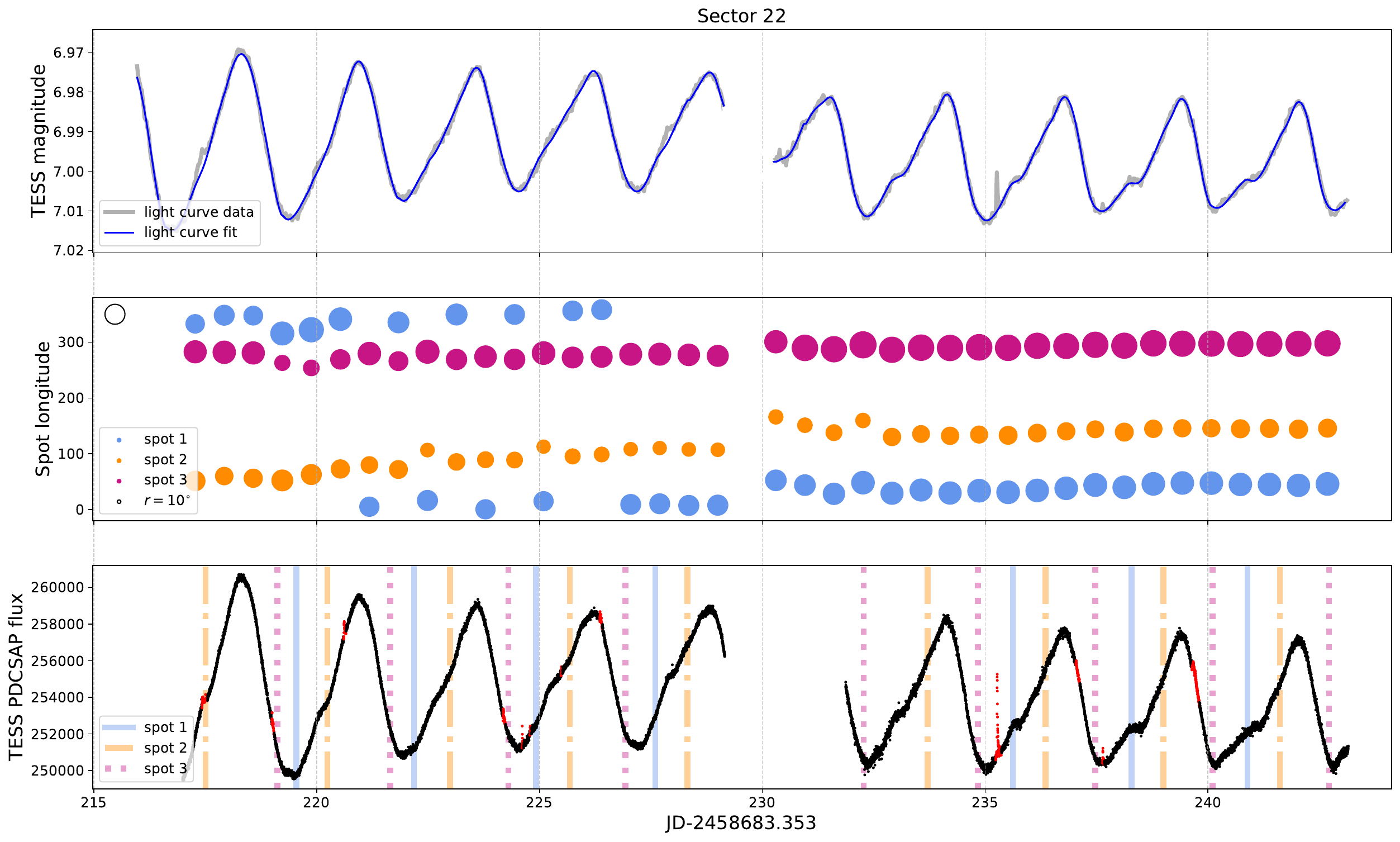}
\end{tabular}
\caption{Same as Fig.\,\ref{fig:sector14} but for TESS sectors S15, S16, S21, S22.}
\label{fig:S15-22}
\end{sidewaysfigure*}

\begin{sidewaysfigure*}
\begin{tabular}{cc}
\includegraphics[width = 0.48\linewidth]{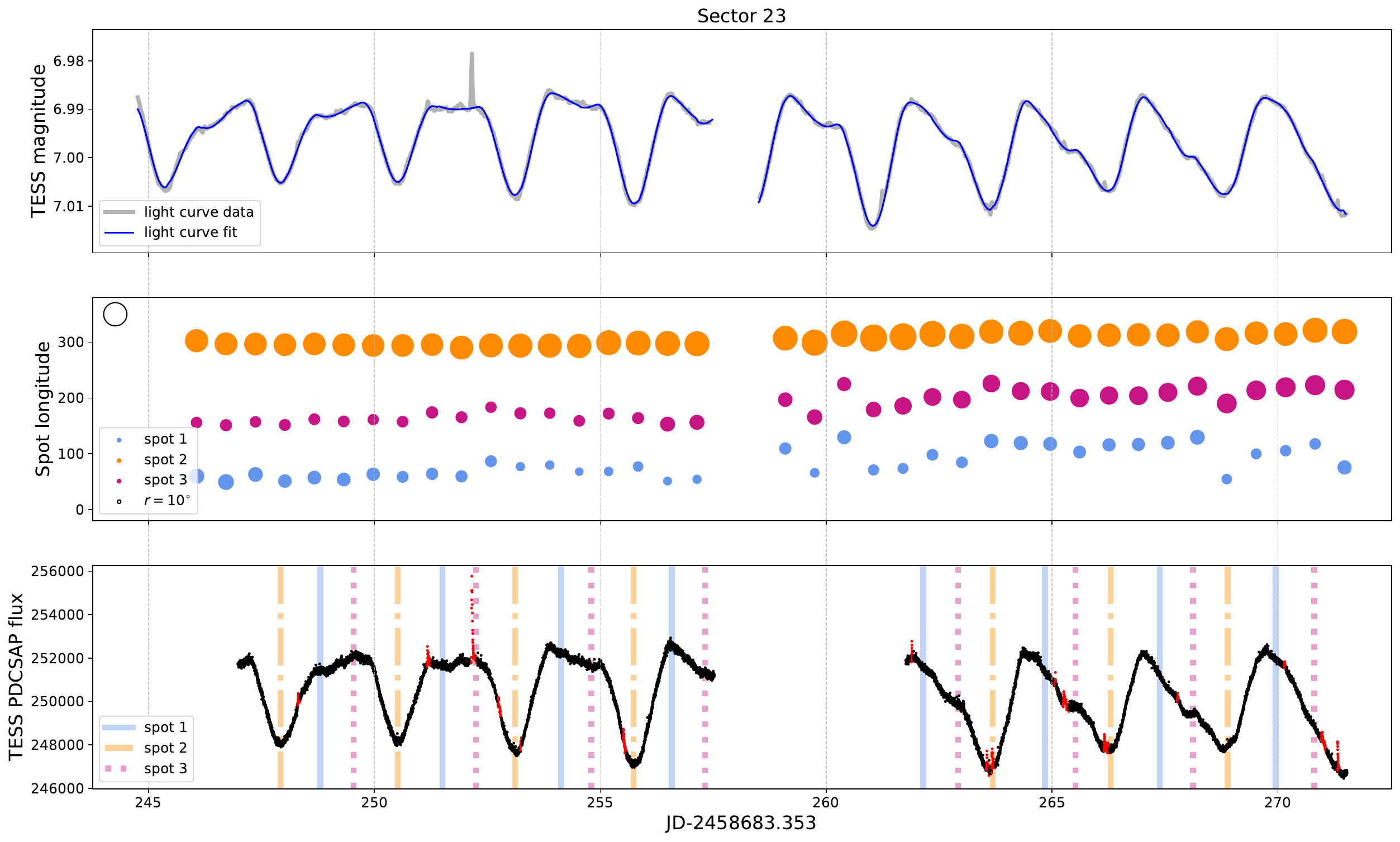} &
\includegraphics[width = 0.48\linewidth]{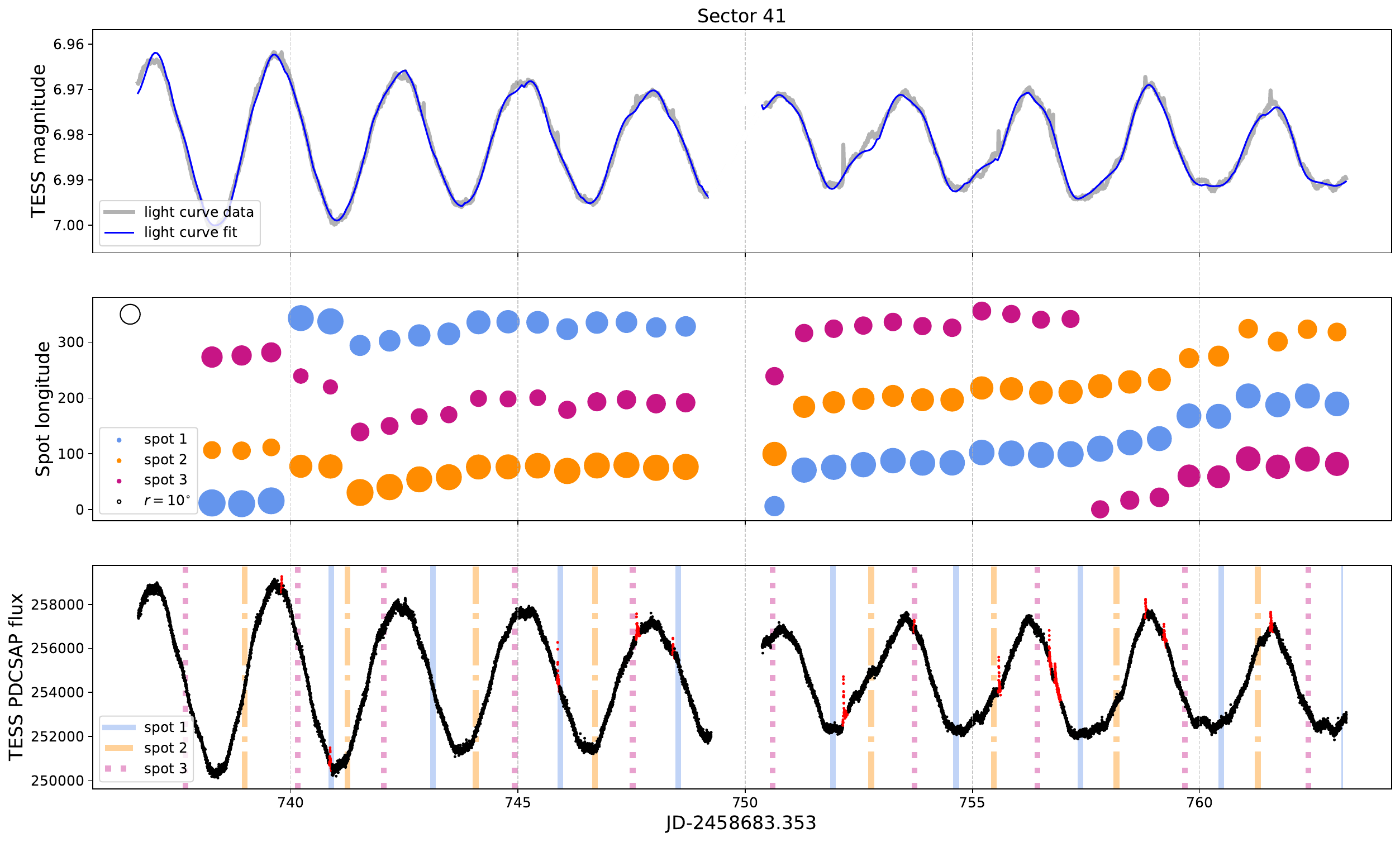} \\
\includegraphics[width = 0.48\linewidth]{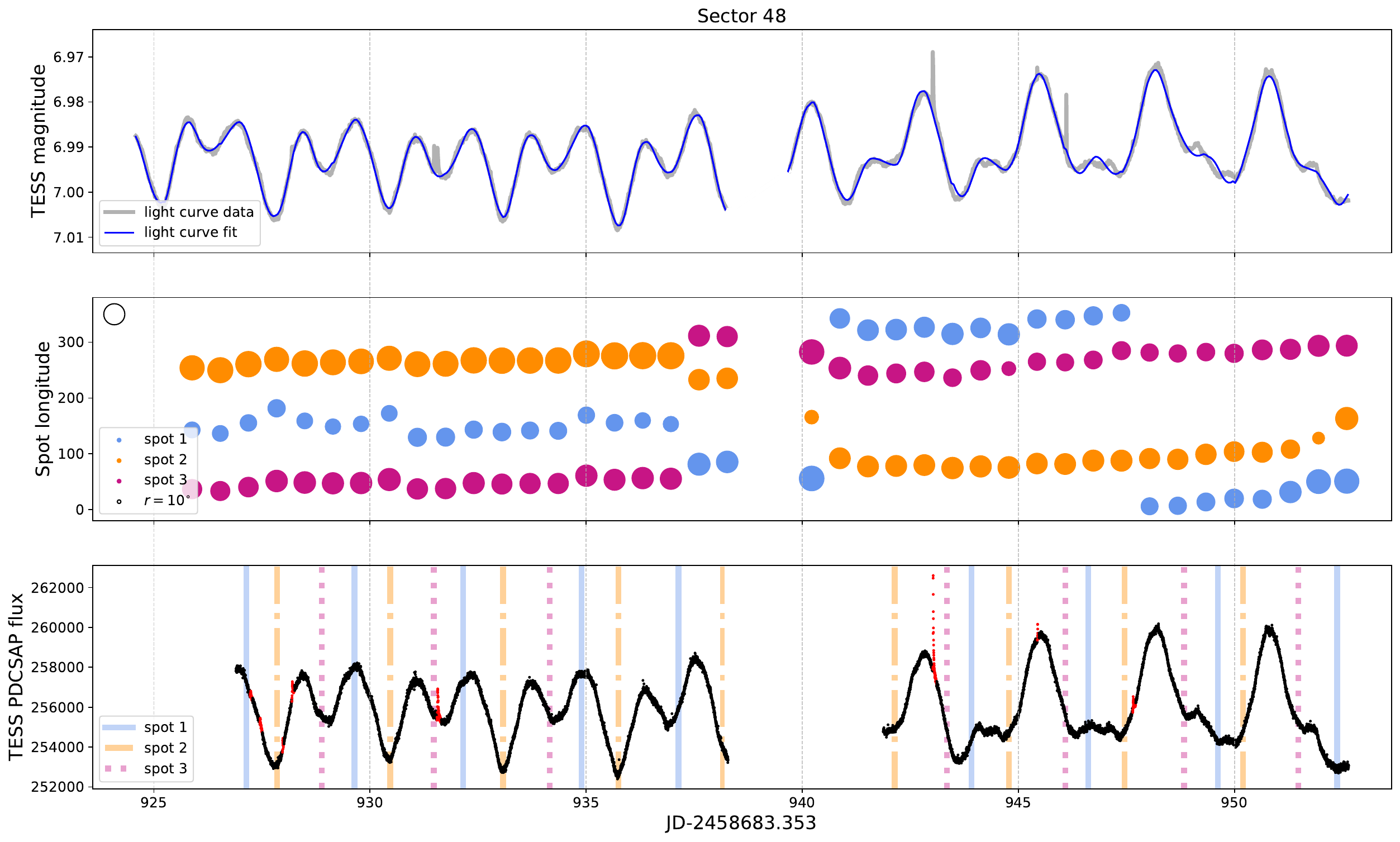} &
\includegraphics[width = 0.48\linewidth]{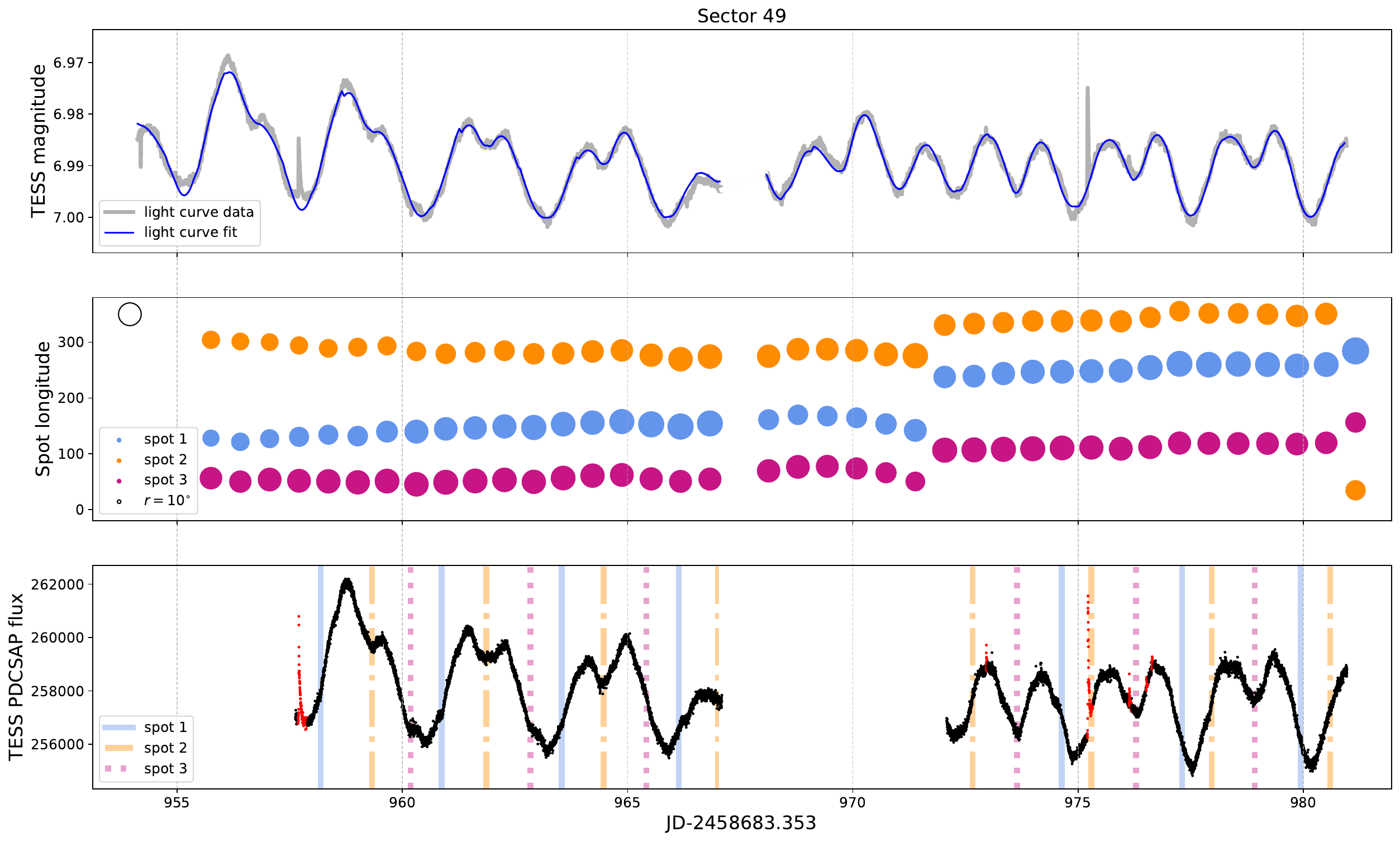}
\end{tabular}
\caption{Same as Fig.\,\ref{fig:sector14} but for TESS sectors S23, S41, S48, S49.}
\label{fig:S23-49}
\end{sidewaysfigure*}

\section{Line profile fits and phase distribution}\label{appC}

\begin{figure*}
\centering
\includegraphics[width=0.9\linewidth]{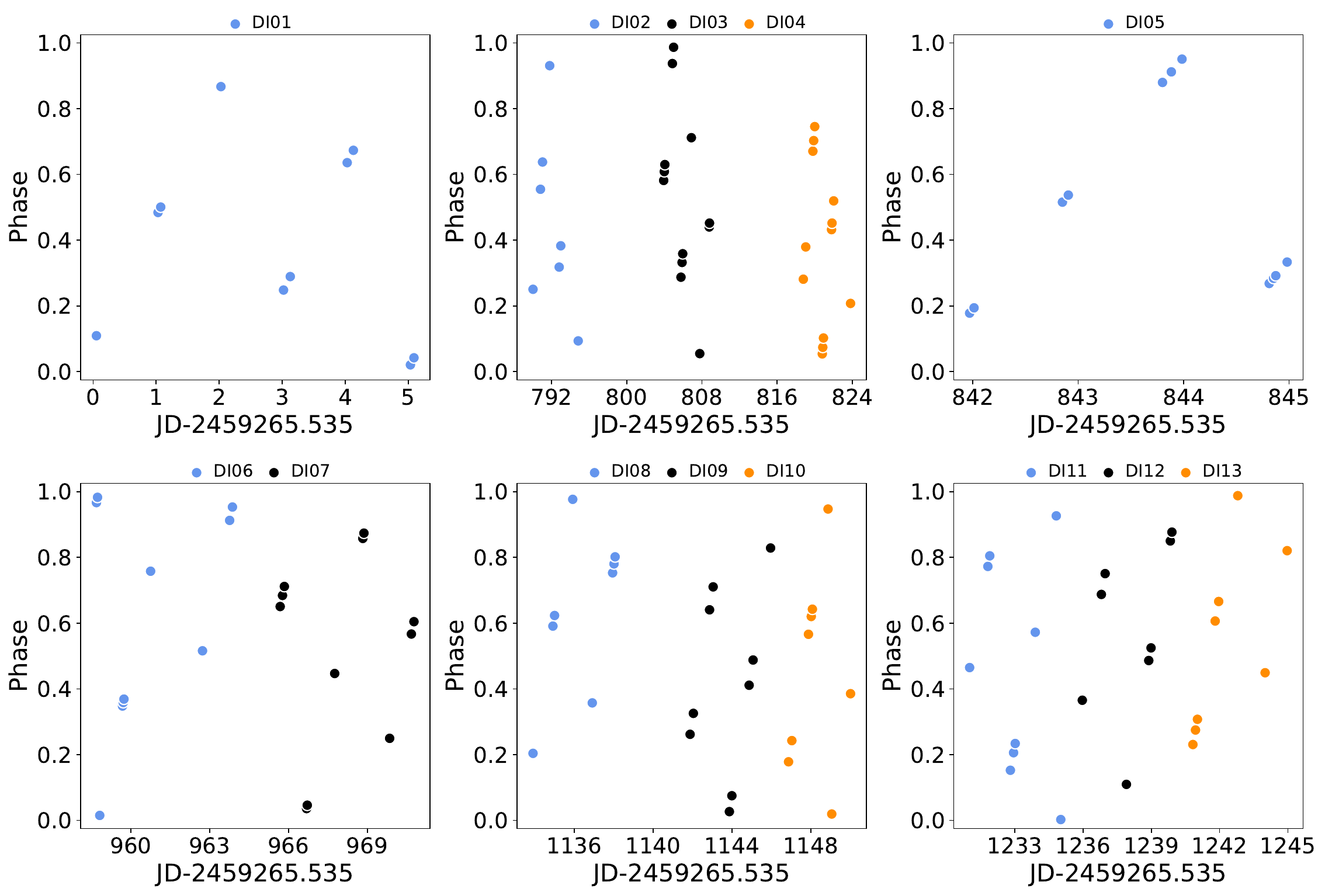}
\caption{Phase distribution-optimized compilation of data subsets DI01-DI13 for Doppler imaging; cf. Table\,\ref{tab:DI_info}. 
}
\label{fig:phases_for_di}
\end{figure*}

\begin{figure*}
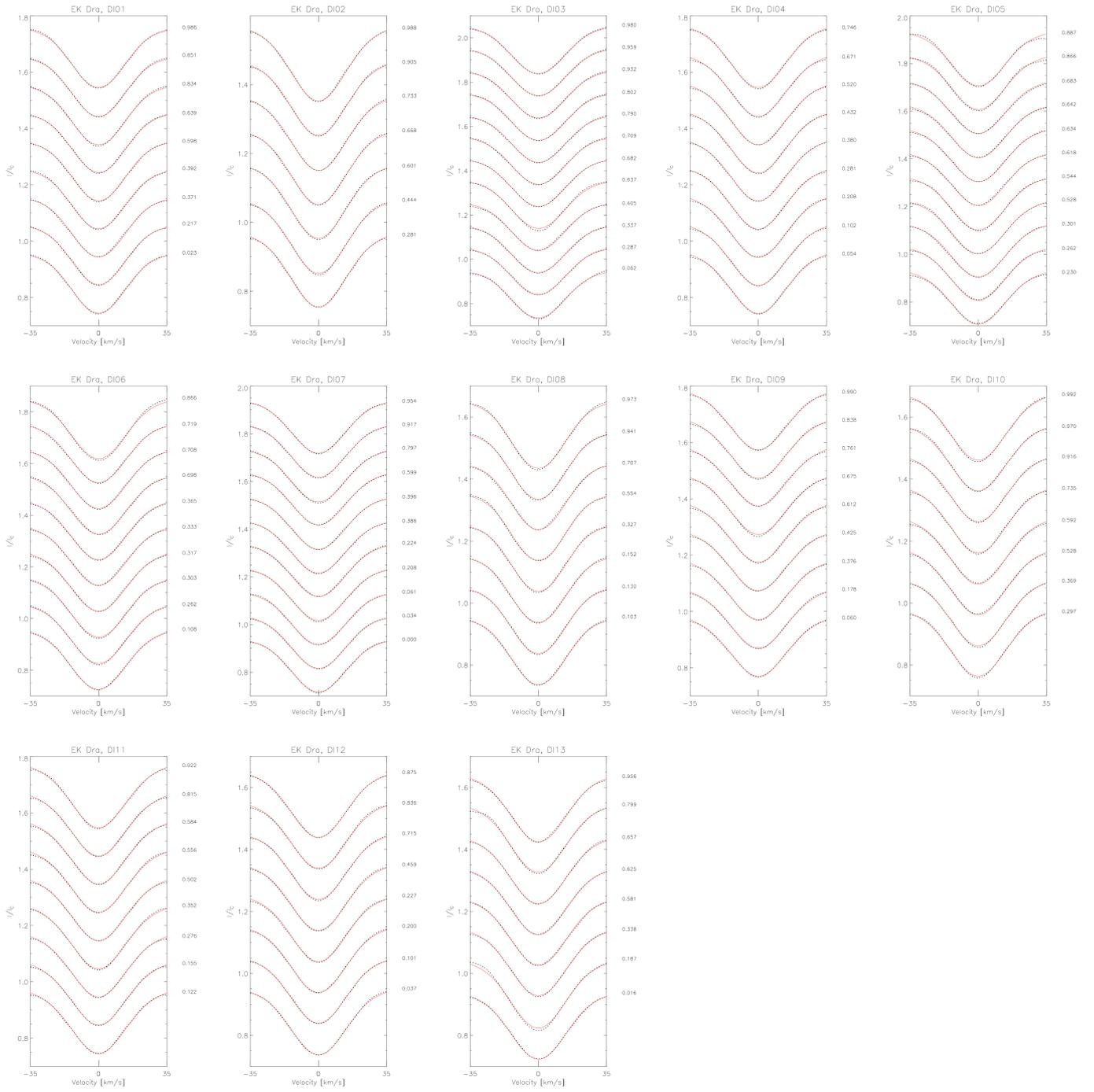

\begin{tabular}{c}
\includegraphics[width = \linewidth]{plots/lineprof1.pdf} \\
\includegraphics[width = \linewidth]{plots/lineprof2.pdf} \\
\includegraphics[width = \linewidth]{plots/lineprof3.pdf} \\
\end{tabular}
\caption{The line profile fits in case of each DI. The black lines are the average line profiles of the imaging lines at a given phase, while the fitted profile is marked in red.}
\label{fig:line_prof}
\end{figure*}

\end{document}